\documentclass[aps,pra,showpacs,twocolumn]{revtex4}
\usepackage{amsmath}
\usepackage{amssymb}
\usepackage{mathrsfs} 

\usepackage{graphicx}
\usepackage{xcolor}
\newcommand\mb{\boldsymbol}

\begin{document}

\title{Self-consistent theory for a plane wave in a moving medium \\ and light-momentum criterion}
\author{Changbiao Wang }
\email{changbiao_wang@yahoo.com}
\affiliation{ShangGang Group, 70 Huntington Road, Apartment 11, New Haven, CT 06512, USA}

\begin{abstract}
A self-consistent theory is developed based on the principle of relativity for a plane wave in a moving non-dispersive, lossless, non-conducting, isotropic, uniform medium.  A light-momentum criterion is set up for the first time, which states that the momentum of light in a medium is parallel to the wave vector in all inertial frames of reference.  By rigorous analysis, novel basic properties of the plane wave are exposed: (a) Poynting vector does not necessarily represent the electromagnetic (EM) power flow when a medium moves; (b) Minkowski light momentum and energy constitute a Lorentz four-vector in a form of single EM-field cell or single photon, and Planck constant is a Lorentz invariant; (c) there is no momentum transfer taking place between the plane wave and the uniform medium, and the EM momentum conservation equation cannot be uniquely determined without resorting to the principle of relativity; and (d) when the medium moves opposite to the wave vector at a faster-than-dielectric light speed, negative frequency and negative EM energy density occur, with the plane wave becoming left-handed.  Finally, a new physics of so-called ``intrinsic Lorentz violation'' is presented as well.
\end{abstract} 

\maketitle 

\section{Introduction}
The momentum of light in a medium is a long-lasting controversial question in physics.  Abraham and Minkowski independently proposed formulations of light momentum. Abraham momentum is inversely proportional to the refractive index of the medium, while Minkowski's is directly proportional to the index.  Experiments claimed to support both Abraham's \cite{r1,r2,r3,r4} and Minkowski's \cite{r5,r6} formulations. Barnett and Loudon assert that the early experiments by Walker \emph{et al.} \cite{r2} ``provide evidence that is no less convincing in favor of the Abraham form'' \cite{r7}, but Feigel insists that ``as far as we know, there are no experimental data that demonstrate the inverse dependence of the radiation pressure on the refractive index'' \cite{r8}; in other words, no experimental observations of light momentum are quantitatively in agreement with the formulation given by Abraham.  The recent direct fiber-recoiling observation by She \emph{et al.} \cite{r4}, which was purported to support the Abraham momentum, is also thought to be ``not uncontroversial'' \cite{r7}. \\
\indent Light momentum has been widely investigated, for all different kinds of dielectric materials, including magnetic \cite{r9,r10} and dispersive \cite{r11} materials, but no agreement has been reached about which formulation is correct.  Comprehensive presentations of the Abraham--Minkowski controversy are given in some review papers \cite{r12,r13,r14}, where there are a lot of valuable references collected.  \\
\indent Maxwell equations support various forms of momentum conservation equations, which is a kind of indeterminacy.  However it is this indeterminacy that results in the question of light momentum.  To find out which formulation of light momentum in a medium is correct, various theories have been proposed.  \vspace{3mm} \\
\indent \textbf{Laue--M\o ller theory.}  Laue and M\o ller proposed a theory where four-vector covariance is imposed on the electromagnetic (EM) energy velocity in a moving medium \cite{r15,r16}.  Laue--M\o ller theory supports Minkowski EM tensor and momentum, because the Minkowski tensor is a real four-tensor while Abraham's is not, as indicated by Veselago and Shchavlev recently \cite{r17}.  But Brevik disagrees, criticizing that such a theory is only ``a test of a tensor's \emph{convenience} rather than its \emph{correctness}'' \cite{r12}.   \vspace{2mm} \\
\indent \textbf{Pfeifer--coworkers theory.}  Pfeifer and coworkers claim that the ``division of the total energy--momentum tensor into electromagnetic and material components is arbitrary'' \cite{r13}.  In other words, the EM part and the material part in the total momentum can be arbitrarily distributed as long as the total momentum is kept the same.  \vspace{2mm}  \\
\indent  \textbf{Mansuripur--Zakharian theory.}  Mansuripur and Zakharian argue that for EM radiation waves, Poynting vector represents EM power flow (energy flow) in any system of materials, and they claim that the Abraham momentum is ``the sole electromagnetic momentum in any system of materials distributed throughout the free space'' \cite{r18}.  \vspace{2 mm} \\
\indent \textbf{Barnett's theory.}  In a recent Letter, Barnett argues that the medium Einstein-box thought experiment (also known as ``Balazs thought experiment'') supports Abraham momentum while the photon--atom Doppler resonance absorption experiment supports Minkowski momentum, and claims that both Abraham and Minkowski momentums are correct: one is kinetic, and the other is canonical \cite{r19}.   \vspace{3mm} \\
\indent Pfeifer--coworkers theory supports ``arbitrary'' EM momentums \cite{r13} while Barnett's theory supports both Abraham and Minkowski momentums \cite{r19}.  Laue--M\o ller theory only supports Minkowski momentum \cite{r15,r16} while Mansuripur--Zakharian theory only supports Abraham momentum \cite{r18}.

Clearly, it is an insufficiency of the Pfeifer--coworkers theory \cite{r13} that the EM momentum in a medium cannot be uniquely determined.  Photons are the carriers of EM momentum for radiation EM waves.  According to Pfeifer--coworkers theory, the momentum of a specific photon in a medium could be Abraham's, Minkowski's, or even arbitrary; thus leading to the momentum not having a determinate value. 

In Barnett's theory \cite{r19}, the argument for supporting Abraham momentum is based on the analysis of the Einstein-box ~thought ~experiment ~by ~the ~``center-of-mass-energy'' approach, where the global momentum--energy conservation law is employed to obtain Abraham photon momentum and energy in the medium box in laboratory frame \cite{r7}.  At first sight, such an approach is indeed impeccable; however, upon more careful investigation, one may find that the approach itself has implicitly assumed the Abraham momentum to be the correct momentum; thus leaving readers an open question: Do the Abraham momentum and energy obtained still satisfy the global momentum--energy conservation law in all inertial frames of reference so that the argument is consistent with the principle of relativity?

Laue--M\o ller theory imposes four-vector covariance on the EM energy velocity in a moving medium, where the energy velocity is defined as the Poynting vector divided by EM energy density \cite{r15,r16}.  Obviously, the Poynting vector is assumed to be the EM power flow in the moving medium.  In Mansuripur--Zakharian theory, the Poynting vector is also assumed to be the EM power flow in any system of materials \cite{r18}.  The two theories have the same basic assumption, but they result in completely different physical conclusions: Minkowski momentum is the unique momentum for Laue--M\o ller theory, while Abraham momentum is the unique momentum for Mansuripur--Zakharian theory.  From this, one may have every reason to question the justification of the assumption used in their theories: Does the Poynting vector really represent the EM power flow in any system of materials, including the moving medium?

In fact, there is another interesting question in Laue--M\o ller theory.  The Laue--M\o ller theory assumes the Poynting vector as the EM power flow (energy flow).  Because the photon is the carrier of the EM energy and momentum, the Minkowski momentum which the theory solely supports is supposed to be parallel to the Poynting vector.  However, the Minkowski momentum and Poynting vector are not parallel in general in a moving medium (see Eqs.\ (\ref{eq37}) and (\ref{eq38}) herein); resulting in a serious contradiction between the basic assumption and conclusion.  

From the preceding analysis, we can see that there are flaws in the existing theories.  Thus the crux of the matter is to set up a self-consistent theory.  This theory must be based on a most fundamental postulate, which constitutes an additional condition imposed on physical laws, so that the light momentum can be uniquely determined.  Such a postulate is the principle of relativity: the laws of physics are the same in all inertial frames of reference \cite{r20}.  This principle is a restriction but also is a guide in formulating physical theories.  According to this principle, there is \emph{no preferred} inertial frame for descriptions of physical phenomena.  For example, Maxwell equations, global momentum and energy conservation laws, Fermat's principle, and Einstein light-quantum hypothesis are equally valid in all inertial frames, no matter whether the medium is moving or at rest, and no matter whether the space is fully or partially filled with a medium.  

In this paper, a self-consistent theory is developed based on the principle of relativity for a plane wave in a moving non-dispersive, lossless, non-conducting, isotropic, uniform medium, which can uniquely determine the light momentum.  In this physical model, photons are introduced by the Einstein light-quantum hypothesis; however, the motion of photons is treated classically through the principle of Fermat, namely all photons are assumed to propagate along light rays.  By analysis of the plane wave, important unconventional conclusions are obtained, which are as follows.

\begin{itemize}
\item There may be a pseudo-power flow when a medium moves, and the Poynting vector does not necessarily denote the EM power flow.  This conclusion explains why the Laue--M\o ller and Mansuripur--Zakharian theories use the same assumption but result in different physical results.
\item Minkowski light momentum and energy constitute a Lorentz four-vector in the form of a single photon or single EM-field cell, and the Planck constant is a Lorentz invariant.  This conclusion has been applied to analysis of the Einstein-box thought experiment, revealing why the argument for Abraham momentum in Barnett's theory is not consistent with the principle of relativity \cite{r21}.
\item There is no momentum transfer taking place between the plane wave and the uniform medium, and the EM momentum conservation equation cannot be uniquely determined without resorting to the principle of relativity.  This conclusion is also supported by the Einstein-box thought experiment analyzed using the EM boundary-condition matching approach, where the leading and trailing light pulse edges in a medium box do not produce additional Lorentz force, and both Abraham and Minkowski momentums satisfy the EM boundary conditions on the vacuum-medium interface \cite{r21}.
\item When the medium moves opposite to the wave vector at a faster-than-dielectric light speed, negative frequency and negative EM energy density occur, with the plane wave becoming left-handed.  In such a case, Minkowski light momentum, Poynting vector (= EM power flow), and the wave vector have the same direction, while the phase velocity is opposite to the Poynting vector because the frequency is negative.
\end{itemize} 

It should be noted that the application of the relativity principle is very tricky, not just manipulating Lorentz transformations.  For example, when applying this principle to the Maxwell equations in free space, one may directly obtain the constancy of light speed \cite{r22}; when applying it to analysis of the Abraham photon momentum in the Einstein-box thought experiment, one may find that the Abraham momentum must have exactly the same form in all inertial frames \cite{r21}; both without any need of Lorentz transformations. \\
\indent According to the principle of relativity, the phase function for a plane wave (see Eq.\ (\ref{eq5})) has the same form in all inertial frames.  From this we can directly obtain an important light-momentum criterion: 
\begin{itemize} 
\item The momentum of light in a medium (including empty space) is parallel to the wave vector in all inertial frames. 
\end{itemize}  
The argument for this criterion is as follows. \\
\indent From the Einstein light-quantum hypothesis, photons are the carriers of light momentum and energy.  Thus the direction of motion of photons is the propagation direction of the light momentum and energy.  The phase function defines equiphase planes of motion (wavefronts), with the wave vector as the normal vector.  From one equiphase plane to another equiphase plane, the path parallel to the normal vector is the shortest and has the minimum optical length.  According to Fermat's principle, light follows the path of least time, and this path is an actual light ray.  Thus the direction of motion of the photons must be parallel to the wave vector, and so must the light momentum.  Because the phase function is invariant in form, this property of light momentum must be valid in all inertial frames.  \\
\indent In some literature, the momentum of light in a medium is defined as the total momentum, namely, the sum of the EM and mechanical parts \cite{r23}.  In this paper, the light momentum is defined as the single photon momentum or EM momentum.  According to this definition, the single photon momentum is the direct result of Einstein light-quantized EM momentum.

\begin{figure}[!ht]  
\begin{center}
\includegraphics[trim=1.2in 6.4in 1.3in 1.1in, clip=true,scale=0.5]{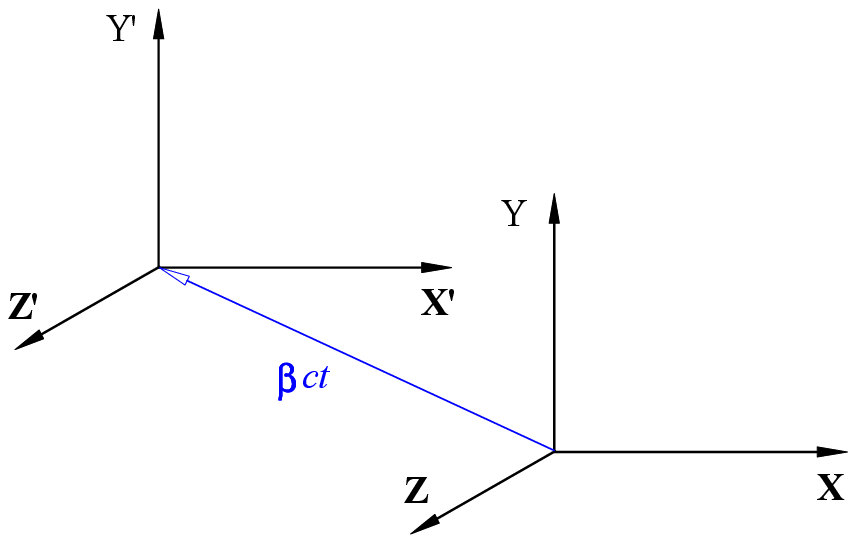}
\end{center}
\caption{Two inertial frames of relative motion.  $X'Y'Z'$ moves with respect to $XYZ$  at  $\mb{\beta}c$, while $XYZ$ moves with respect to $X'Y'Z'$ at $\mb{\beta}'c$  (not shown), with  $\mb{\beta}'=-\mb{\beta}$.  Note:  $(\gamma\mb{\beta},\gamma)$  is the four-vector describing the motion of  $X'Y'Z'$, while $(\gamma'\mb{\beta}',\gamma')$  with $\gamma'=\gamma$  is the four-vector describing the motion of  $XYZ$; thus $(\gamma\mb{\beta},\gamma)$  and $(\gamma'\mb{\beta}',\gamma')$  are not the same four-vector, which is an exception in this primed-unprimed symbol usage.  }
\label{fig1}
\end{figure}
  
\indent It should be emphasized that the principle of relativity is the backbone of the theory developed in the present paper.  One may insist that the medium should define a \emph{preferred} inertial frame of reference so that there is \emph{no} reason why Fermat's principle is valid in all inertial frames.  For example, Ravndal suggested a \emph{preferred} Lorentz transformation when a dielectric medium exists \cite{r24}.  However in this paper, the principle of relativity is taken as a fundamental postulate whether with or without the existence of a medium, and the standard Lorentz transformation \cite{r20} is assumed to be universal. \\
\indent The paper is organized as follows.  In Sec.\ II, refractive index, phase velocity (photon velocity), and group velocity are defined for a plane wave in a moving uniform medium.  In Sec.\ III, single photon momentum is analyzed.  In Sec.\ IV, novel basic properties of a plane wave are revealed.  Finally in Sec.\ V, some conclusions and remarks are given.

\section{Refractive Index, Phase Velocity, and Group Velocity}
In this section, invariant forms of refractive index, phase velocity (photon velocity), and group velocity are defined for a plane wave in a moving uniform medium.  An unconventional analysis of the relation between the group velocity and Poynting vector is given.

Suppose that the frame $X'Y'Z'$ moves with respect to the laboratory frame $XYZ$ at a constant velocity of $\mb{\beta}c$, with all corresponding coordinate axes in the same directions and their origins overlapping at $t=t'=0$, as shown in Fig. \ref{fig1}.  The Lorentz transformation of the time--space four-vector $(\mathbf{x},ct)$ is given by \cite{r25} \\
\begin{align}
&\mathbf{x}=\mathbf{x}'+\frac{\gamma-1}{\beta^2}(\mb{\beta}'\cdot\mathbf{x}')\mb{\beta}'-\gamma\mb{\beta}'ct',
\label{eq1}
\\
&ct=\gamma(ct'-\mb{\beta}'\cdot\mathbf{x}'),
\label{eq2}
\end{align} \\
where $c$ is the universal light speed, and $\gamma=(1-\mb{\beta}^2)^{-1/2}$  is the time dilation factor.  

The EM fields $\mathbf{E}$ and $\mathbf{B}$, and $\mathbf{D}$ and $\mathbf{H}$ respectively constitute a covariant second-rank anti-symmetric tensor $F^{\alpha\beta}(\mathbf{E,B})$  and  $G^{\alpha\beta}(\mathbf{D,H})$, of which the Lorentz transformations can be written in intuitive three-dimensional vector forms, given by \cite{r25,r26} \\
\begin{equation}
\left[\hspace{-0.2em}\begin{array}{c} \mathbf{E} \\ \mathbf{D} \end{array}\hspace{-0.2em}\hspace{-0.2em}\right]=\gamma\left[\hspace{-0.2em}\begin{array}{c} \mathbf{E}' \\ \mathbf{D}' \end{array}\hspace{-0.2em}\right]+\gamma\mb{\beta}'\times\left[\hspace{-0.2em}\begin{array}{c} \mathbf{B}'c \\ \mathbf{H}'/c \end{array}\hspace{-0.2em}\right]-\frac{\gamma-1}{\beta^2}\mb{\beta}'\cdot\left[\hspace{-0.2em}\begin{array}{c} \mathbf{E}' \\ \mathbf{D}' \end{array}\hspace{-0.2em}\right]\mb{\beta}',
\label{eq3}
\end{equation} 

\begin{equation}
\left[\hspace{-0.2em}\begin{array}{c} \mathbf{B} \\ \mathbf{H} \end{array}\hspace{-0.2em}\right]=\gamma\left[\hspace{-0.2em}\begin{array}{c} \mathbf{B}' \\ \mathbf{H}' \end{array}\hspace{-0.2em}\right]-\gamma\mb{\beta}'\times\left[\hspace{-0.2em}\begin{array}{c} \mathbf{E}'/c \\ \mathbf{D}'c \end{array}\hspace{-0.2em}\right]-\frac{\gamma-1}{\beta^2}\mb{\beta}'\cdot\left[\hspace{-0.2em}\begin{array}{c} \mathbf{B}' \\ \mathbf{H}' \end{array}\hspace{-0.2em}\right]\mb{\beta}',
\label{eq4}
\end{equation} \\
\noindent with  $\mathbf{E\cdot B}$, $\mathbf{E}^2-(\mathbf{B}c)^2$, $\mathbf{D\cdot H}$,  $(\mathbf{D}c)^2-\mathbf{H}^2$, and  $\mathbf{E}\cdot\mathbf{D}-\mathbf{B}\cdot\mathbf{H}$ as Lorentz invariants \cite{r27}.

\subsection{Refractive index and its Lorentz \\transformation}
Suppose that there is a plane wave propagating in the medium-rest frame  $X'Y'Z'$, and the plane wave has a phase function given by  $\Psi'(\mathbf{x}',t')=\omega't'-n'_d\mathbf{k}'\cdot\mathbf{x}'$, where  $\omega' ~(>0)$ is the angular frequency, $n'_d\mathbf{k}'$ is the wave vector, $n'_d\equiv |n'_d\mathbf{k}'|/|\omega'/c|$  is the refractive index of the medium, and $|\mathbf{k}'|=\omega'/c$.  It is seen from Eqs.\ (\ref{eq3}) and (\ref{eq4}) that the phase function $\Psi(\mathbf{x},t)$  for this plane wave observed in the laboratory frame $XYZ$ must be equal to $\Psi'(\mathbf{x}',t')$  (see Sec.\ IV), namely, invariance of phase.  Thus we have \\
\begin{equation}
\Psi=\omega t-n_d\mathbf{k}\cdot\mathbf{x}=\omega't'-n'_d\mathbf{k}'\cdot\mathbf{x}',
\label{eq5}
\end{equation} \\
where $n_d\mathbf{k}$   is the wave vector in the laboratory frame, $n_d\equiv |n_d\mathbf{k}|/|\omega/c|$  is the refractive index, and $|\mathbf{k}|=|\omega/c|$.  Note that $\omega$  can be negative \cite{r28}. 

From the covariance of $(\mathbf{x}',ct')$  and the invariance of the phase, we conclude that $(n'_d\mathbf{k}',\omega'/c)$  must be Lorentz covariant \cite{r20}.  By setting the time--space four-vector $X^{\mu} = (\mathbf{x},ct)$  and the wave four-vector $K^{\mu}=(n_d\mathbf{k},\omega/c)$, Eq.\ (\ref{eq5}) can be written in a covariant form, given by $(\omega t-n_d\mathbf{k}\cdot\mathbf{x})=g_{\mu\nu}K^{\mu}X^{\nu}$  with the metric tensor  $g_{\mu\nu}=g^{\mu\nu}=diag(-1,-1,-1,+1)$ \cite{r29}.  Because $X^{\mu}$ must fulfill four-vector Lorentz rule, the phase invariance and the covariance of $K^{\mu}$ are equivalent.  

From Eqs.\ (\ref{eq1}) and (\ref{eq2}) with $n'_d\mathbf{k}'\rightarrow\mathbf{x}'$ and $\omega'/c\rightarrow ct'$, we obtain $K^{\mu}=(n_d\mathbf{k},\omega/c)$.  Setting $\mathbf{\hat{n}}'=n'_d\mathbf{k}'/|n'_d\mathbf{k}'|$  as the unit wave vector we have \\
\begin{equation}
\omega=\omega'\gamma(1-n'_d\mathbf{\hat{n}}'\cdot\mb{\beta}'), \hspace{1cm} \mbox{(Doppler formula)}
\label{eq6}
\end{equation}
\begin{equation}
n_d\mathbf{k}=(n'_d\mathbf{k}')+\frac{\gamma-1}{\beta^2}(n'_d\mathbf{k}')\cdot\mb{\beta}'\mb{\beta}'-\gamma\mb{\beta}'\left( \frac{\omega'}{c} \right).
\label{eq7}
\end{equation} \\

\indent Because $K_{\mu}K^{\mu}=g^{\mu\nu}K_{\mu}K_{\nu}$  is a Lorentz scalar, we have \\
\begin{equation}
g^{\mu\nu}(K_{\mu}K_{\nu}-K'_{\mu}K'_{\nu})=0,
\label{eq8}
\end{equation}
namely, 
\begin{equation}
\left( \frac{\omega}{c} \right)^2-(n_d\mathbf{k})^2=\left( \frac{\omega'}{c} \right)^2-(n'_d\mathbf{k}')^2,
\label{eq9}
\end{equation} \\
which indicates that $\omega^2(1-n_d^2)=\omega'^2(1-n'^2_d)$  is a Lorentz invariant; thus we have $n'_d>1\Rightarrow n_d>1$.  From Eqs.\ (\ref{eq6}) and (\ref{eq9}) we obtain the Lorentz transformation of the refractive index, given by \\
\begin{equation}
n_d=\frac{\sqrt{(n'^2_d-1)+\gamma^2(1-n'_d\mathbf{\hat{n}}'\cdot\mb{\beta}')^2}}{|\gamma(1-n'_d\mathbf{\hat{n}}'\cdot\mb{\beta}')|},
\label{eq10}
\end{equation} \\
from which we can see that the motion of the dielectric medium results in an anisotropic refractive index.  But in free space with  $n'_d=1$, we have $n_d=1$  holding for any propagation directions of waves, namely, empty space is always isotropic.

\subsection{Phase velocity and photon \\propagation velocity}
It is seen from Eq.\ (\ref{eq5}) that the phase function is symmetric with respect to all inertial frames, independent of which frame the medium is fixed in; accordingly, no frame should make its phase function have any priority in time and space.  From this we can conclude that the definitions of equiphase plane and phase velocity should be symmetric, independent of the choice of inertial frames.  Thus the phase velocity can be defined as \\
\begin{equation}
\mb{\beta}_{ph}c=\frac{\omega}{|n_d\mathbf{k}|}\mathbf{\hat{n}}=\frac{c}{n_d}\frac{\omega}{|\omega|}\mathbf{\hat{n}}=\beta_{ph}c\hspace{0.15em}\mathbf{\hat{n}},
\label{eq11}
\end{equation} \\
leading to 
\begin{equation}
\omega-n_d\mathbf{k}\cdot\mb{\beta}_{ph}c=0,
\label{eq12}
\end{equation} \\
where $\mathbf{\hat{n}}=n_d\mathbf{k}/|n_d\mathbf{k}|$ is the unit wave vector in the laboratory frame, and $\mb{\beta}_{ph}c$  and $K^{\mu}$  are related through $K^{\mu}=(n_d\mathbf{k},\omega/c)=\omega(n_d/c)^2(\mb{\beta}_{ph}c,c/n_d^2)$.  Note that the definition of the phase velocity $\mb{\beta}_{ph}c$ is based on the wave four-vector $K^{\mu}$, while the velocity definition of a \emph{massive particle} is based on the time--space four-vector  $X^{\mu}$.  Because the phase velocity $\mb{\beta}_{ph}c$  is parallel to  $n_d\mathbf{k}$, which is a constraint, there is no ``phase velocity four-vector''.

One may conjecture that $\gamma_{ph}(\mb{\beta}_{ph}c,c)$  could be the ``phase velocity four-vector'', with  $\gamma_{ph}=(1-\beta_{ph}^2)^{-1/2}$; however, by further examination one can find that it is not true because $\gamma_{ph}(\mb{\beta}_{ph}c,c)$ does not fulfill the four-vector Lorentz rule.

\begin{figure*}[!ht]   
\begin{center}
\includegraphics[trim=1.2in 6.9in 1.2in 1.0in, clip=true,scale=0.75]{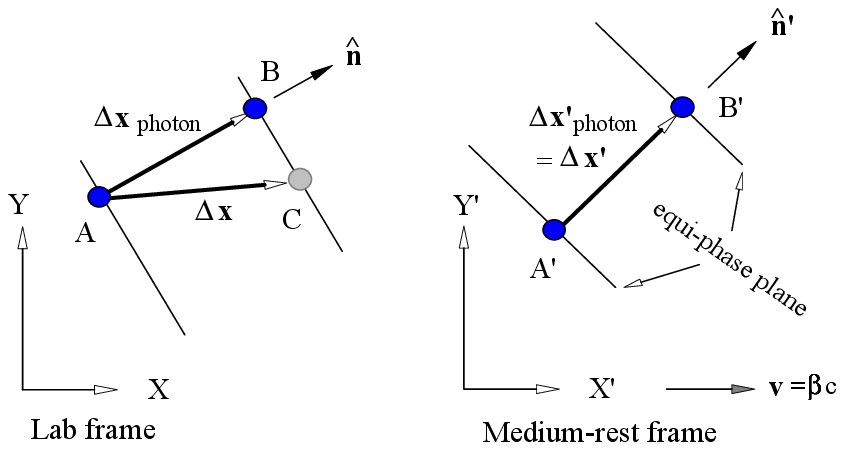}
\end{center}
\caption{Photon real and apparent displacements. The photon propagation velocity is the phase velocity.  From Fermat's principle and the principle of relativity, when a photon together with its associated equiphase plane moves from $A'$   to $B'$  along the unit wave vector $\mathbf{\hat{n}}'$  in the medium rest-frame, it moves from $A$ to $B$ observed in the lab frame.  However, because the time--space coordinates may not reflect its real location, resulting in an illusion, the photon appears to have moved to $C$ in terms of the time--space Lorentz transformation.  Thus the photon \emph{real} displacement $\Delta\mathbf{x}_{photon}$  only can be converted from its \emph{apparent} displacement $\Delta\mathbf{x}$  through  $\Delta\mathbf{x}_{photon}=(\mathbf{\hat{n}}\cdot\Delta\mathbf{x})\mathbf{\hat{n}}$, and the phase velocity is related through $\mb{\beta}_{ph}c=d\mathbf{x}_{photon}/dt=(\mathbf{\hat{n}}\cdot\mathbf{u})\mathbf{\hat{n}}$, where $\mathbf{u}\equiv d\mathbf{x}/dt$  is the photon \emph{apparent velocity}, with $\mathbf{u}'=(c/n'_d)\mathbf{\hat{n}}'$ in the medium-rest frame.  Note that $(\mb{\beta}_{ph}c)\|\Delta\mathbf{x}_{photon}$  and  $\mathbf{u}\|\Delta\mathbf{x}$.   }
\label{fig2}
\end{figure*}

From Eq.\ (\ref{eq5}), the equiphase-plane (wavefront) equation of motion is given by  $\omega t-|n_d\mathbf{k}|\mathbf{\hat{n}}\cdot\mathbf{x}=const$, with $\mathbf{\hat{n}}$ as the unit normal vector of the plane, leading to $\omega-|n_d\mathbf{k}|\mathbf{\hat{n}}\cdot(d\mathbf{x}/dt)=0$.  Comparing with Eq.\ (\ref{eq11}), we obtain $\mb{\beta}_{ph}c=\mathbf{\hat{n}}\cdot(d\mathbf{x}/dt)\mathbf{\hat{n}}$.  Thus we have a physical explanation for  $\mb{\beta}_{ph}c$: the phase velocity is equal to the changing rate of the equiphase plane's \emph{distance} displacement $\mathbf{\hat{n}}(\mathbf{\hat{n}}\cdot d\mathbf{x})$ over time  $dt$, and it is the photon propagation velocity.  Obviously, this photon-velocity definition is consistent with Fermat's principle in all inertial frames: Light follows the path of least time.  

In general, $d\mathbf{x}/dt$  in the expression $\mb{\beta}_{ph}c=\mathbf{\hat{n}}\cdot(d\mathbf{x}/dt)\mathbf{\hat{n}}$  is undetermined unless a definition is given.  If $d\mathbf{x}'/dt'=\mb{\beta}_{ph}'c$  is assigned in the \emph{medium-rest} frame, we call $\mathbf{u}\equiv d\mathbf{x}/dt$  the photon \emph{apparent} velocity (``apparent'' here means ``looks like but is not necessarily real'').  Note that $\gamma_u(\mathbf{u},c)$  with $\gamma_u=(1-\mathbf{u}^2/c^2)^{-1/2}$  is a four-vector.  Thus we have  $\mb{\beta}_{ph}c=(\mathbf{\hat{n}}\cdot\mathbf{u})\mathbf{\hat{n}}$, with $|\mb{\beta}_{ph}c|\leq|\mathbf{u}|$, and $|\mb{\beta}_{ph}c|=|\mathbf{u}|$  if $\mathbf{u}\|\mathbf{\hat{n}}$.

It can be shown that the photon apparent velocity,  $\mathbf{u}$, Poynting vector,  $\mathbf{S}=\mathbf{E}\times\mathbf{H}$, and EM energy density, $W_{em}=0.5(\mathbf{D}\cdot\mathbf{E}+\mathbf{B}\cdot\mathbf{H})$, are related through  $\mathbf{u}=\mathbf{S}/W_{em}$, where $\mathbf{S}/W_{em}$  is traditionally the so-called ``energy velocity'' \cite{r16}.  Calculations indicate  $|\mathbf{u}|=(1-\xi)^{1/2}c\leq c$, where $\xi=(n'^2_d-1)/[\gamma^2(n'_d-\mathbf{\hat{n}}'\cdot\mb{\beta}')^2]\geq 0$, and thus we have $|\mb{\beta}_{ph}c|\leq|\mathbf{u}|\leq c$, as expected.

From the equiphase-plane equation $\omega t-n_d\mathbf{k}\cdot\mathbf{x}=const$ $\Rightarrow$ $\omega-n_d\mathbf{k}\cdot\mathbf{u}=0$ $\Rightarrow$ $\mb{\beta}_{ph}c=(\mathbf{\hat{n}}\cdot\mathbf{u})\mathbf{\hat{n}}$, we have introduced the photon apparent velocity, $\mathbf{u}$.  The appearance of $\mathbf{u}$ comes from: the real photon velocity is the phase velocity $\mb{\beta}_{ph}c$, which is defined based on the wave four-vector $K^{\mu}$ instead of the time--space four-vector  $X^{\mu}$.  From this it follows that, when using the time--space coordinates to describe the motion of a photon, the space coordinates may not reflect the photon real location, resulting in an illusion.  Thus there must be a conversion between the photon apparent and real locations.  This conversion is governed by the photon real versus apparent velocity equation, $\mb{\beta}_{ph}c=(\mathbf{\hat{n}}\cdot\mathbf{u})\mathbf{\hat{n}}$, from which we have  $\Delta\mathbf{x}_{photon}=(\mathbf{\hat{n}}\cdot\Delta\mathbf{x})\mathbf{\hat{n}}$, where $\Delta\mathbf{x}_{photon}\equiv \mb{\beta}_{ph}c\Delta t$  is the photon \emph{real} displacement and $\Delta\mathbf{x}\equiv\mathbf{u}\Delta t$  is its \emph{apparent} displacement, as shown in Fig. \ref{fig2}.  Note that $(\Delta\mathbf{x},c\Delta t)$  is a four-vector while $(\Delta\mathbf{x}_{photon},c\Delta t)$  is not, except for in free space ($n_d=1$ ), where the ``empty space'' is isotropic, and the Poynting vector $\mathbf{S}=W_{em}\mathbf{u}$ is always parallel to the wave vector in all inertial frames, leading to $\mb{\beta}_{ph}c=\mathbf{u}=c\mathbf{\hat{n}}$  and  $\Delta\mathbf{x}_{photon}=\mb{\beta}_{ph}c\Delta t=\mathbf{u}\Delta t=\Delta\mathbf{x}$.

Now let us check the conservation law of photon Minkowski angular momentum.  Photon momentum is given by $\hbar n_d\mathbf{k}$  (see Eq.\ (\ref{eq16})).  Without loss of generality, suppose that the photon is located at $\mathbf{x}=\mathbf{x}'=0$  when $t=t'=0$.  Thus we have $\mathbf{x}_{photon}=\Delta\mathbf{x}_{photon}=(\mathbf{\hat{n}}\cdot\Delta\mathbf{x})\mathbf{\hat{n}}
=(\mathbf{\hat{n}}\cdot\mathbf{x})\mathbf{\hat{n}}$, and $\mathbf{x}_{photon}\times\hbar n_d\mathbf{k}=0$, namely the photon angular momentum is conserved in all inertial frames.

\subsection{Group velocity and its relation with \\Poynting vector}
The classical definition of group velocity is given by $\mathbf{v}_{gr-c}=\partial\omega/\partial(n_d\mathbf{k})$, defined in the normal direction to the wave-vector surface \cite{r26,r30}.  In this paper, we suggest a \emph{modified} definition, given by $\mathbf{v}_{gr}=\mathbf{\hat{n}}\partial\omega / \partial|n_d\mathbf{k}|$, defined in the wave-vector direction.  Obviously, $\mathbf{v}_{gr}\cdot\mathbf{\hat{n}}=\mathbf{v}_{gr-c}\cdot\mathbf{\hat{n}}$ holds between the two definitions.  

From Eq.\ (\ref{eq5}), we know that the form-invariant definition of refractive index $n_d=|n_d\mathbf{k}|/|\omega/c|$ itself also defines a dispersion equation of $(n_d\mathbf{k})^2-(\mathbf{n}_d\omega/c)^2=0$ for the plane wave, where $\mathbf{n}_d=n_d\mathbf{k}/(\omega/c)$ with $n_d=|\mathbf{n}_d|$ is the refractive-index vector \cite{r30}.  From the Maxwell equations $\mathbf{B}c=\mathbf{n}_d\times\mathbf{E}$ and $\mathbf{D}c=-\mathbf{n}_d\times\mathbf{H}$ (see Eq.\ (\ref{eq26})), we have $\check{\epsilon}\cdot\mathbf{E}c^2+\mathbf{n}_d\times[~\check{\mu}^{-1}\cdot(\mathbf{n}_d\times\mathbf{E})]=0$, which is a system of linear equations for $(E_x,E_y,E_z)$, and where $\check{\epsilon}$ and $\check{\mu}$ are the dielectric permittivity and permeability tensors, respectively, and $\check{\mu}^{-1}$ denotes the inverse tensor of $\check{\mu}$.  From this, we obtain the (eigen) Fresnel equation $F(n_d,\epsilon_{ij},\mu_{ij},\theta_w,\phi_w)=0$ \cite{r30}, or $n_d=n_d(\epsilon_{ij},\mu_{ij},\theta_w,\phi_w)$, with $\epsilon_{ij}$ and $\mu_{ij}$ being the dielectric tensor elements, and $\theta_w$ and $\phi_w$ the wave-vector angles so that $n_{d}k_x=|n_d\mathbf{k}|\sin\theta_w\cos\phi_w$,  $n_{d}k_y=|n_d\mathbf{k}|\sin\theta_w\sin\phi_w$, and $n_{d}k_z=|n_d\mathbf{k}|\cos\theta_w$. Note that $n_d$ does not explicitly contain $|n_d\mathbf{k}|$.  If there is any dispersion,  $n_d$ implicitly contains $\omega$ through $\epsilon_{ij}$ and $\mu_{ij}$.  Thus from the modified group-velocity definition $\mathbf{v}_{gr}=\mathbf{\hat{n}}\partial\omega/\partial|n_d\mathbf{k}|$, we obtain \\
\begin{equation}
\textbf{v}_{gr}=\frac{\mb{\beta}_{ph}c}{1+(\omega/n_d)(\partial n_d/\partial\omega)}.
\label{eq13}
\end{equation} \\
Because the dielectric medium is assumed to be non-dispersive for the physical model considered in the present paper, $\partial n_d/\partial\omega=0$ is valid.  Thus we have $\mathbf{v}_{gr}=\mb{\beta}_{ph}c$, namely, the group velocity is equal to the phase velocity, and parallel to the wave vector.  

As we know, for a plane wave in an anisotropic medium the wave vector and Poynting vector usually are not parallel.  It has been thought that the group velocity is parallel to the Poynting vector, instead of the wave vector, as shown in the classical electrodynamics textbook by Landau and Lifshitz \cite{r30}.

The moving isotropic medium becomes an anisotropic medium, as seen in Eq.\ (\ref{eq10}); however, the group velocity we obtained is  $\mathbf{v}_{gr}=\mb{\beta}_{ph}c$, parallel to the wave vector instead of the Poynting vector.  Obviously, this is not in agreement with the result in the textbook \cite{r30}.  

Why do we have to modify the group velocity definition?  By careful analysis we find that there is some flaw in the classical definition, for which an argument is given as follows.  

Following the Landau--Lifshitz approach in analysis of a plane wave in an anisotropic lossless medium \cite{r30}, with the holding of $(\delta\mathbf{D}\cdot\mathbf{E}-\mathbf{D}\cdot\delta\mathbf{E})
+(\delta\mathbf{B}\cdot\mathbf{H}-\mathbf{B}\cdot\delta\mathbf{H})=0$  taken into account for a moving \emph{non-dispersive} uniform medium, from Maxwell equations we obtain  $\delta\omega=\mathbf{S}\cdot\delta(n_d\mathbf{k})/W_{em}$, where $\delta(n_d\mathbf{k})$  is an arbitrary infinitesimal change in wave vector, $\mathbf{S}=\mathbf{E}\times\mathbf{H}$ is the Poynting vector, and $W_{em}=0.5(\mathbf{E}\cdot\mathbf{D}+\mathbf{B}\cdot\mathbf{H})$ is the EM energy density.  From the mathematical definition of the gradient  $\partial\omega/\partial(n_d\mathbf{k})=\mathbf{v}_{gr-c}$, we have $\delta\omega=\mathbf{v}_{gr-c}\cdot\delta(n_d\mathbf{k})$  for an arbitrary $\delta(n_d\mathbf{k})$.  Comparing   $\delta\omega=\mathbf{S}\cdot\delta(n_d\mathbf{k})/W_{em}$ and  $\delta\omega=\mathbf{v}_{gr-c}\cdot\delta(n_d\mathbf{k})$, we have  $\mathbf{v}_{gr-c}=\mathbf{S}/W_{em}$, namely the classical group velocity is equal to the ``energy velocity'' \cite{r16}, parallel to the Poynting vector.  (Note: $\check{\epsilon}$ and $\check{\mu}$  in $\mathbf{D}=\check{\epsilon}\cdot\mathbf{E}$ and $\mathbf{B}=\check{\mu}\cdot\mathbf{H}$ are not symmetric in general for a \emph{moving} medium so that $\delta\mathbf{D}\cdot\mathbf{E}-\mathbf{D}\cdot\delta\mathbf{E}=0$  and $\delta\mathbf{B}\cdot\mathbf{H}-\mathbf{B}\cdot\delta\mathbf{H}=0$ cannot separately hold, unlike in the traditional anisotropic medium where the symmetry of $\check{\epsilon}$ and $\check{\mu}$  is assumed \cite{r30}.)

However, there is a serious flaw for $\mathbf{v}_{gr-c}=\mathbf{S}/W_{em}$, because $|\mathbf{v}_{gr-c}|$  can be greater than the phase velocity  $|\mb{\beta}_{ph}c|$, as shown in Sec.\ IV, which is not physical for a non-dispersive lossless medium.  The modified definition $\mathbf{v}_{gr}=\mathbf{\hat{n}}\partial\omega / \partial|n_d\mathbf{k}|$, which leads to $\mathbf{v}_{gr}=\mb{\beta}_{ph}c$ for a non-dispersive medium, has removed this flaw, which is the reason why the classical definition of group velocity must be modified.

Because the modified group velocity, Eq.\ (\ref{eq13}), is always parallel to the wave vector $n_d\mathbf{k}$ instead of the Poynting vector, the Poynting vector does not necessarily denote the direction of power flow; this is clearly confirmed by the strict EM field solutions given in Sec.\ IV (see Eq.\ (\ref{eq38})).

\section{Four-vector covariance of Minkowski photon momentum and energy}
In this section, single photon momentum in a medium is analyzed based on Einstein light-quantum hypothesis, and it is shown that the Minkowski photon momentum is strongly supported by Lorentz four-vector covariance, and it meets light-momentum criterion.  The Fizeau running water experiment is reanalyzed as a support to the Minkowski momentum.  

For a uniform plane wave, observed in the medium-rest frame the EM fields  $\mathbf{E}'$,  $\mathbf{B}'$,  $\mathbf{D}'$, and $\mathbf{H}'$  are related through $\mathbf{B}'c/n'_d=\mathbf{\hat{n}}'\times\mathbf{E}'$  and $\mathbf{H}'=\mathbf{\hat{n}}'\times(c/n'_d)\mathbf{D}'$ (consult Sec.\ IV).  Thus the Minkowski and Abraham EM momentum density vectors can be expressed as \\
\begin{equation}
\mathbf{g}'_M=\mathbf{D}'\times\mathbf{B}'=\frac{n'_d}{c}(\mathbf{D}'\cdot\mathbf{E}')\mathbf{\hat{n}}',
\label{eq14}
\end{equation}

\begin{equation}
\mathbf{g}'_A=\frac{\mathbf{E}'\times\mathbf{H}'}{c^2}=\frac{1}{n'_d c}(\mathbf{D}'\cdot\mathbf{E}')\mathbf{\hat{n}}'.
\label{eq15}
\end{equation} \\
Note that $(\mathbf{E}'\times\mathbf{H}')\|\mathbf{\hat{n}}'$  holds in the medium-rest frame, but $(\mathbf{E}\times\mathbf{H})\|\mathbf{\hat{n}}$  is not valid in general in the laboratory frame (see Eq.\ (\ref{eq38})).  According to Einstein light-quantum hypothesis, the EM energy density $\mathbf{D}'\cdot\mathbf{E}'$  is proportional to single photon energy $\hbar\omega'$ ($\omega'>0$), namely, $\mathbf{D}'\cdot\mathbf{E}'=N'_p\hbar\omega'$  where $N'_p$ is the photon number density, and the EM momentum density vectors $\mathbf{g}'_M$  and $\mathbf{g}'_A$  are proportional to single photon momentums $\mathbf{p}'_M$  for Minkowski's and $\mathbf{p}'_A$  for Abraham's, respectively, namely, $\mathbf{g}'_M=N'_p\mathbf{p}'_M$  and  $\mathbf{g}'_A=N'_p\mathbf{p}'_A$.  Thus from Eqs.\ (\ref{eq14}) and (\ref{eq15}), we obtain $\mathbf{p}'_M=\mathbf{\hat{n}}'n'_d\hbar\omega'/c$  and  $\mathbf{p}'_A=\mathbf{\hat{n}}'\hbar\omega'/(n'_dc)$.  

The Minkowski photon momentum also can be naturally obtained from the covariance of relativity of wave four-vector, as follows. 

Suppose that the Planck constant $\hbar$  is a Lorentz scalar (see Sec.\ IV).  From the given definition of wave four-vector,    $K'^{\mu}=(n'_d\mathbf{k}',\omega'/c)=(\mathbf{\hat{n}}'n'_d\omega'/c,\omega'/c)$ multiplied by $\hbar$, we obtain a ``momentum--energy four-vector'', which has exactly the same form as a \emph{massive} particle's, given by \\
\begin{equation}
P'^{\mu}=(\hbar n'_d\mathbf{k}',\hbar\omega'/c)=(\mathbf{p}',E'/c),
\label{eq16}
\end{equation} \\
where $E'=\hbar\omega'$  is the photon energy.  In terms of the four-vector structure, $\mathbf{p}'$  must be the momentum; thus we have the photon momentum in a medium, given by $\mathbf{p}'=\mathbf{\hat{n}}'n'_d\hbar\omega'/c$, that is, the Minkowski photon momentum $\mathbf{p}'_M$  obtained from Eq.\ (\ref{eq14}).  

Because $\hbar K'^{\mu}$  is a four-vector, the Minkowski photon momentum $\mathbf{p}'_M=\hbar n'_d\mathbf{k}'$  is parallel to the wave vector in all inertial frames, and thus it meets light-momentum criterion.

On the other hand, because the Minkowski photon momentum and energy constitute a Lorentz four-vector, Abraham's must not, otherwise mathematical contradictions would result, except for in free space where Minkowski and Abraham momentums are identical \cite{r21}.

From the principle of relativity, we have the invariance of phase, from which we have the covariant wave four-vector.  From the wave four-vector combined with the Einstein light-quantum hypothesis, we have the Minkowski photon momentum, which strongly supports the consistency of Minkowski momentum with the relativity and light-momentum criterion. 

In the classical electrodynamics, the Fizeau running water experiment is usually taken to be experimental evidence of the relativistic four-velocity addition rule \cite{r25}.  In fact, it should be taken to be in support of the Minkowski momentum because the photon has no four-velocity.  To better understand this, let us make a simple analysis, as follows.

Suppose that the running-water medium is at rest in the $X'Y'Z'$ frame.  Because the Minkowski momentum--energy $(\hbar n'_d\mathbf{k}', \hbar\omega'/c)$ is four-vector covariant, Eqs.\ (\ref{eq9}) and (\ref{eq10}) hold.  Setting $-\mb{\beta}'=\mb{\beta}=\beta\mathbf{\hat{n}}'$  (the water moves parallel to the wave vector), from Eq.\ (\ref{eq10}) we have the refractive index in the laboratory frame, given by \\
\begin{equation}
n_d=\left| \frac{n'_d+\beta}{1+n'_d\beta} \right|.
\label{eq17}
\end{equation} \\
\indent Thus from Eq.\ (\ref{eq11}), the light speed (= phase velocity = photon velocity) in the running water, observed in the laboratory frame, is given by \\
\begin{equation}
|\mb{\beta}_{ph}c|=\frac{c}{n_d}\approx\frac{c}{n'_d}\left[1+\beta\left(n'_d-\frac{1}{n'_d}\right)\right] 
\label{eq18}
\end{equation} \\
for $|\beta|\ll1$, which is the very formula confirmed by the Fizeau experiment.  When the water runs along (opposite to) the wave vector direction, we have $\beta>0$  $(\beta<0)$  and the light speed is increased (reduced).

One might argue for a different photon energy in a medium.  In the medium-rest frame, the dispersion equation, directly resulting from the second-order wave equation, is given by \cite{r26}
\begin{equation}
(n'_d\omega'/c)^2-(n'_d\mathbf{k}')^2=0,
\label{eq19}
\end{equation} 

\noindent which actually is the expression, in the medium-rest frame, of the general definition of the form-invariant refractive index $n_d=|n_d\mathbf{k}|/|\omega/c|$.  This dispersion relation was thought to be a characterization of the relation between EM energy and momentum, and the photon energy in a medium was suggested to be $n'_d\hbar\omega'$ to keep a zero rest energy (see \S3.1a of Ref. \cite{r26}, for example).  However, it should be noted that, although Eq.\ (\ref{eq19}) is Lorentz invariant in form, $(n'_d\mathbf{k}',n'_d\omega'/c)$ is not a Lorentz covariant four-vector, because only $K'^{\mu}=(n'_d\mathbf{k}',\omega'/c)$ is; except for $n'_d=1$.  If using $(n'_d\hbar\mathbf{k}',n'_d\hbar\omega'/c)$  to define the photon momentum--energy four-vector, then it is not Lorentz covariant.

Thus it is justifiable to define $\hbar K'^{\mu}=(n'_d\hbar\mathbf{k}', \hbar\omega'/c)$  as the photon momentum--energy four-vector, as is done in Eq.\ (\ref{eq16}), because $\hbar K'^{\mu}$  is four-vector covariant, with Eq.\ (\ref{eq19}) as a natural result.  

\section{Novel properties of a plane Wave in a moving medium}
A plane wave is the simplest strict solution to the Maxwell equations \cite{r26}; however, its physics is far from being well understood.  In this section, we will explore novel basic properties for a plane wave in a moving non-dispersive, lossless, non-conducting, isotropic, uniform medium.  Specifically, we will show that (a) the Poynting vector does not necessarily represent EM power flow when a medium moves; (b) Minkowski EM momentum and energy constitute a Lorentz four-vector, and the Planck constant is a Lorentz invariant; (c) there is no momentum transfer between the plane wave and medium, and the EM momentum conservation equation cannot be uniquely determined without resorting to the principle of relativity; and (d) when the medium moves opposite to the wave vector at a faster-than-dielectric light speed, negative frequency and negative EM energy density result, with the plane wave becoming left-handed.

Suppose that the plane-wave solution in the medium-rest frame $X'Y'Z'$ is given by  \\
\begin{equation}
(\mathbf{E}', \mathbf{B}', \mathbf{D}', \mathbf{H}')=(\mathbf{E}'_0, \mathbf{B}'_0, \mathbf{D}'_0, \mathbf{H}'_0)\cos\Psi',
\label{eq20}
\end{equation} \\
\noindent where  $\Psi'=(\omega't'-n'_d\mathbf{k}'\cdot\mathbf{x}')$, with  $\omega'>0$; and $(\mathbf{E}'_0, \mathbf{B}'_0, \mathbf{D}'_0, \mathbf{H}'_0)$ are real constant amplitude vectors.  $\mathbf{D}'=\epsilon'\mathbf{E}'$  and $\mathbf{B}'=\mu'\mathbf{H}'$ hold, where $\epsilon'>0$ and $\mu'>0$  are the constant dielectric permittivity and permeability, respectively. As required by wave equation, the refractive index $n'_d\equiv |n'_d\mathbf{k}'|/|\omega'/c|$ is given by $n'_d=c\sqrt{\epsilon'\mu'}$, where $n'_d\ge 1$ is assumed to hold.  Thus $(\mathbf{E}', \mathbf{B}', \mathbf{\hat{n}}')$  and $(\mathbf{D}', \mathbf{H}', \mathbf{\hat{n}}')$  are, respectively, two sets of right-hand orthogonal vectors, with $\mathbf{E}'=(c/n'_d)\mathbf{B}'\times\mathbf{\hat{n}}'$ and $\mathbf{H}'=\mathbf{\hat{n}}'\times(c/n'_d)\mathbf{D}'$, and $\mathbf{E}'\times\mathbf{H}'=(\mathbf{D}'\cdot\mathbf{E}')(c/n'_d)\textbf{\^n}'=(\mathbf{D}'\cdot\mathbf{E}')(\mb{\beta}'_{ph}c)$, resulting from the Maxwell equations. 

Inserting Eq.\ (\ref{eq20}) into Eqs.\ (\ref{eq3}) and (\ref{eq4}), we obtain the plane-wave solution in the laboratory frame  $XYZ$, given by  \\
\begin{equation}
(\mathbf{E}, \mathbf{B}, \mathbf{D}, \mathbf{H})=(\mathbf{E}_0, \mathbf{B}_0, \mathbf{D}_0, \mathbf{H}_0)\cos\Psi,
\label{eq21}
\end{equation} \\
\noindent where $\Psi=(\omega t-n_d\mathbf{k}\cdot\mathbf{x})$, and the phase factor $\cos\Psi=\cos\Psi'$ must hold for any time--space points $\Rightarrow\Psi(\mathbf{x},t)=\Psi'(\mathbf{x}',t')+2l\pi$ with $l$ an integer, but $\Psi=\Psi'=0$ holds when $\mathbf{x}=\mathbf{x}'=0$ and  $t=t'=0$, $\Rightarrow l=0$, or $\Psi=\Psi'$, namely, ``invariance of phase''; $(\mathbf{E}_0, \mathbf{B}_0, \mathbf{D}_0, \mathbf{H}_0)$ are given by 
\begin{align}
\left[ \begin{array}{c} \mathbf{E}_0 \\ \mathbf{H}_0 \end{array} \right]=\gamma(1&-n'_d\mathbf{\hat{n}}'\cdot\mb{\beta}')\left[ \begin{array}{c} \mathbf{E}'_0 \\ \mathbf{H}'_0 \end{array} \right]  \nonumber
 \\    
&+~( \gamma n'_d\mathbf{\hat{n}}'-\frac{\gamma-1}{\beta^2}\mb{\beta}')
\left[ \begin{array}{c} \mb{\beta}'\cdot\mathbf{E}'_0 \\ \mb{\beta}'\cdot\mathbf{H}'_0 \end{array} \right],
\label{eq22}
\end{align}
\begin{align}
\left[ \begin{array}{c} \mathbf{B}_0 \\ \mathbf{D}_0 \end{array} \right]=\gamma(1&-\frac{1}{n'_d}\mathbf{\hat{n}}'\cdot\mb{\beta}')\left[ \begin{array}{c} \mathbf{B}'_0 \\ \mathbf{D}'_0 \end{array} \right]  \nonumber
\\    
&+~( \gamma \frac{1}{n'_d}\mathbf{\hat{n}}'-\frac{\gamma-1}{\beta^2}\mb{\beta}')
\left[ \begin{array}{c} \mb{\beta}'\cdot\mathbf{B}_0' \\ \mb{\beta}'\cdot\mathbf{D}'_0 \end{array} \right].
\label{eq23}
\end{align}  \\
\indent Note that the transformations in Eqs.\ (\ref{eq22}) and (\ref{eq23}) are ``synchronous''; for example, $\mathbf{E}_0$  is expressed only in terms of $\mathbf{E}'_0$.  All field quantities have the same phase factor, whether in the medium-rest frame or laboratory frame.  It is clearly seen from Eq.\,(\ref{eq20})--Eq.\,(\ref{eq23}) that the invariance of phase,  $\Psi=\Psi'$, is a natural result.  In the following analysis, formulas are derived in the laboratory frame, because they are invariant in form in all inertial frames.  

Under Lorentz transformations, the Maxwell equations keep the same forms as in the medium-rest frame, given by 
\begin{align}
&\nabla \times\mathbf{E}=-\partial\mathbf{B}/\partial t,  \hspace{8mm} \nabla\cdot\mathbf{D}=\rho,
\label{eq24}
\\
&\nabla \times\mathbf{H}=\mathbf{J}+\partial\mathbf{D}/\partial t, \hspace{4mm} \nabla\cdot\mathbf{B}=0,
\label{eq25}
\end{align} 
with $\mathbf{J}=0$  and $\rho=0$  for the plane wave.  From these, we have 
\begin{equation}
\omega\mathbf{B}=n_d\mathbf{k}\times\mathbf{E}, \hspace{1mm} \mbox{and}  \hspace{2mm} \omega\mathbf{D}=-n_d\mathbf{k}\times\mathbf{H},
\label{eq26}
\end{equation}
leading to  $\mathbf{D}\cdot\mathbf{E}=\mathbf{B}\cdot\mathbf{H}$, namely, the electric energy density is equal to the magnetic energy density, which is valid in all inertial frames.

From Eqs.\ (\ref{eq22}) and (\ref{eq23}), by tedious calculations we can obtain intuitive expressions for examining the space relations of EM fields observed in the laboratory frame, given by \\
\begin{align}
&\mathbf{D}\cdot\mathbf{\hat{n}}=\mathbf{D}'\cdot\mathbf{\hat{n}}'=0, \hspace{4mm} \mathbf{B}\cdot\mathbf{\hat{n}}=\mathbf{B}'\cdot\mathbf{\hat{n}}'=0,
\label{eq27}
\\
&\mathbf{E}\cdot\mathbf{\hat{n}}=\frac{\gamma(n'^2_d-1)(\mathbf{E}'\cdot\mb{\beta}')}{\sqrt{(n'^2_d-1)+\gamma^2(1-n'_d\mathbf{\hat{n}}'\cdot\mb{\beta}')^2}},
\label{eq28}
\\
&\mathbf{H}\cdot\mathbf{\hat{n}}=\frac{\gamma(n'^2_d-1)(\mathbf{H}'\cdot\mb{\beta}')}{\sqrt{(n'^2_d-1)+\gamma^2(1-n'_d\mathbf{\hat{n}}'\cdot\mb{\beta}')^2}},
\label{eq29}
\end{align}  
and
\begin{align}
&\mathbf{E}\cdot\mathbf{B}=\mathbf{E}'\cdot\mathbf{B}'=0, \hspace{2mm} \mathbf{D}\cdot\mathbf{H}=\mathbf{D}'\cdot\mathbf{H}'=0,~~~~~~~~~
\label{eq30}
\\
&\mathbf{E}\cdot\mathbf{H}=\gamma^2(n'^2_d-1)(\mb{\beta}'\cdot\mathbf{E}')(\mb{\beta}'\cdot\mathbf{H}'),~~~~~~~~~~~~~~~~
\label{eq31}
\\
&\mathbf{D}\cdot\mathbf{B}=-\gamma^2\left(1-\frac{1}{n'^2_d}\right)(\mb{\beta}'\cdot\mathbf{D}')(\mb{\beta}'\cdot\mathbf{B}'),~~~~~~~~~~
\label{eq32}
\\
&\mathbf{D}\cdot\mathbf{E}=\gamma^2(1-n'_d\mathbf{\hat{n}}'\cdot\mb{\beta}')\left(1-\frac{1}{n'_d}\mathbf{\hat{n}}'\cdot\mb{\beta}'\right)\mathbf{D}'\cdot\mathbf{E}',
\label{eq33}
\\
&\mathbf{B}\cdot\mathbf{H}=\gamma^2(1-n'_d\mathbf{\hat{n}}'\cdot\mb{\beta}')\left(1-\frac{1}{n'_d}\mathbf{\hat{n}}'\cdot\mb{\beta}'\right)\mathbf{B}'\cdot\mathbf{H}',
\label{eq34}
\end{align} 
and from Eq.\ (\ref{eq26}) we have \\
\begin{align}
&\mathbf{E}=(\mathbf{\hat{n}}\cdot\mathbf{E})\mathbf{\hat{n}}-\beta_{ph}c\hspace{0.15em}\mathbf{\hat{n}}\times\mathbf{B},
\label{eq35}
\\
&\mathbf{H}=(\mathbf{\hat{n}}\cdot\mathbf{H})\mathbf{\hat{n}}+\beta_{ph}c\hspace{0.15em}\mathbf{\hat{n}}\times\mathbf{D}.
\label{eq36}
\end{align} \\
\indent It can be seen from the preceding equations that  $\mathbf{E}\bot\mathbf{B}$,  $\mathbf{B}\bot\mathbf{\hat{n}}$, $\mathbf{D}\bot\mathbf{H}$, and $\mathbf{D}\bot\mathbf{\hat{n}}$ hold in the laboratory frame, but  $\mathbf{E}\|\mathbf{D}$,  $\mathbf{B}\|\mathbf{H}$,  $\mathbf{E}\bot\mathbf{H}$,  $\mathbf{D}\bot\mathbf{B}$,  $\mathbf{E}\bot\mathbf{\hat{n}}$, and $\mathbf{H}\bot\mathbf{\hat{n}}$  usually do not hold any more. 

\subsection{Pseudo-power flow due to \\ motion of the medium}
As seen in Eq.\ (\ref{eq10}), a moving isotropic uniform medium becomes an anisotropic medium, and as a result, a pseudo-power flow may be incurred, which is shown as follows.  

From Eqs.\ (\ref{eq26}), (\ref{eq35}), and (\ref{eq36}), we obtain Minkowski EM momentum and the Poynting vector, given by  \\
\begin{align}
&\mathbf{D}\times\mathbf{B}=\left(\frac{\mathbf{D}\cdot\mathbf{E}}{\omega}\right)n_d\mathbf{k}=\left(\frac{n_d}{c}\right)^2(\mathbf{D}\cdot\mathbf{E})\mathbf{v}_{gr},
\label{eq37}
\\
&\mathbf{E}\times\mathbf{H}=v_{gr}[v_{gr}(\mathbf{D}\times\mathbf{B})-(\mathbf{\hat{n}}\cdot\mathbf{H})\mathbf{B}-(\mathbf{\hat{n}}\cdot\mathbf{E})\mathbf{D}],
\label{eq38}
\end{align}  \\
where  $\mathbf{v}_{gr}$, with $v_{gr}=\mathbf{v}_{gr}\cdot\mathbf{\hat{n}}$ and $|v_{gr}|=c/n_d$, is the group velocity obtained from Eq.\ (\ref{eq13}) with no dispersion ($\partial n_d/\partial\omega=0$) considered, which is equal to the phase velocity $\mb{\beta}_{ph}c$, as defined by Eq.\ (\ref{eq11}).  Note that the Minkowski momentum $\mathbf{D}\times\mathbf{B}$  has the same direction as the wave vector $n_d\mathbf{k}$, while the Poynting vector $\mathbf{E}\times\mathbf{H}$  has three components: one in  the $(\mathbf{D}\times\mathbf{B})$-direction, one in the $\mathbf{B}$-direction, and one in the $\mathbf{D}$-direction; the latter two are perpendicular to the group velocity $\mathbf{v}_{gr}=\mb{\beta}_{ph}c$  or the wave vector $n_d\mathbf{k}$.

We can divide the Poynting vector $\mathbf{S}=\mathbf{E}\times\mathbf{H}$ into two parts, namely, $\mathbf{S}=\mathbf{S}_{power}+\mathbf{S}_{pseu}$, where  \\
\begin{align}
&\mathbf{S}_{power}=v_{gr}^2(\mathbf{D}\times\mathbf{B})=(\mathbf{D}\cdot\mathbf{E})\mathbf{v}_{gr}=W_{em}\mathbf{v}_{gr},
\label{eq39}
\\
&\mathbf{S}_{pseu}=-v_{gr}[(\mathbf{\hat{n}}\cdot\mathbf{H})\mathbf{B}+(\mathbf{\hat{n}}\cdot\mathbf{E})\mathbf{D}],
\label{eq40}
\end{align} \\
with $\mathbf{S}_{power}$ and $\mathbf{S}_{pseu}$ being perpendicular to each other ($\mathbf{S}_{power}\bot\mathbf{S}_{pseu}$ ).  

From Eqs.\ (\ref{eq38}), (\ref{eq39}), and (\ref{eq37}) with Eqs. (\ref{eq6}) and (\ref{eq33}) taken into account, we have $(n_d\mathbf{k})\cdot (\mathbf{E}\times\mathbf{H})=(n_d\mathbf{k})\cdot\mathbf{S}_{power}=\omega (\mathbf{D}\cdot\mathbf{E})>0$ holding in the sense of excluding those discrete zero points where $\mathbf{D}\cdot\mathbf{E}=(\mathbf{D}_0\cdot\mathbf{E}_0)\cos^2(\omega t-n_d\mathbf{k}\cdot\mathbf{x})=0$.  From Eqs.\ (\ref{eq11}), (\ref{eq38}), and (\ref{eq37}), we have $(\mb{\beta}_{ph}c)\cdot(\mathbf{E}\times\mathbf{H})=(c/n_d)^2 (\mathbf{D}\cdot\mathbf{E})$.

From Eq.\ (\ref{eq39}), we find that $\mathbf{S}_{power}$ carries all the EM energy $W_{em}=0.5(\mathbf{D}\cdot\mathbf{E}+\mathbf{B}\cdot\mathbf{H})$ moving at the group velocity $\mathbf{v}_{gr}$, and it is a real power flow.  According to the energy conservation law, $\mathbf{S}_{pseu}$ should not be responsible for any EM energy transport, and it is a pseudo-power flow.  Thus the energy velocity, defined as $\mathbf{S}_{power}/W_{em}$, is equal to the group velocity, $\mathbf{v}_{gr}$, and the phase velocity, $\mb{\beta}_{ph}c$, which is justifiable when considering that the medium is assumed to be non-dispersive and lossless. 

In the medium-rest frame, both $\mathbf{E}'\bot (n'_d\mathbf{k}')$ and $\mathbf{H}'\bot(n'_d\mathbf{k}')$ hold.  Thus if $\mb{\beta}'\|(n'_d\mathbf{k}')$ holds, we have $\mathbf{E}'\cdot\mb{\beta}'=0\Rightarrow\mathbf{E}\cdot\mathbf{\hat{n}}=0$ from Eq.\ (\ref{eq28}), and $\mathbf{H}'\cdot\mb{\beta}'=0\Rightarrow\mathbf{H}\cdot\mathbf{\hat{n}}=0$ from Eq.\ (\ref{eq29}).  From this, according to Eq.\ (\ref{eq40}) we find that the pseudo-power flow $\mathbf{S}_{pseu}$ vanishes only when the medium ($n'_d\ne 1$) moves at $\mb{\beta}c=-\mb{\beta}'c$ parallel to the wave vector $n'_d\mathbf{k}'$.  In other words, $\mathbf{S}_{pseu}\ne 0$ is incurred in general in a \emph{moving medium} ($n'_d\ne 1$ ), so that the Poynting vector $\mathbf{E}\times\mathbf{H}$ is not parallel to the wave vector $n_d\mathbf{k}$.

The physical difference between $\mathbf{S}_{power}$ and $\mathbf{S}_{pseu}$ also can be seen from divergence theorem.  The divergence of  $\mathbf{S}_{power}$  is given by \\
\begin{align}
\nabla\cdot&\mathbf{S}_{power}=-\frac{\partial W_{em}}{\partial t} \nonumber
\\
&=-(\mathbf{D}_0\cdot\mathbf{E}_0)\frac{\partial}{\partial t}\cos^2\ (\omega t-n_d\mathbf{k}\cdot\mathbf{x})\ne 0
\label{eq41}
\end{align} \\
holding except for those discrete points, which means that $\mathbf{S}_{power}$ is responsible for an EM power flowing into and out of ~a differential box, ~but~ the time average $<\nabla\cdot\mathbf{S}_{power}>=0$, meaning that the powers going in and out are the same on time average, with no net energy left in the box.  In contrast, $\nabla\cdot\mathbf{S}_{pseu}\equiv 0$  holds resulting from  $\nabla\cdot\mathbf{B}=0$, $\nabla\cdot\mathbf{D}=0$,  $\mathbf{B}\bot(n_d\mathbf{k})$, and $\mathbf{D}\bot(n_d\mathbf{k})$, which means that $\mathbf{S}_{pseu}$ is not responsible for a power flowing at any time for any places (otherwise energy conservation would be broken).

Because $\mathbf{S}_{power}$ and $\mathbf{S}_{pseu}$  are perpendicular to each other, $|\mathbf{S}/W_{em}|>|\mathbf{S}_{power}/W_{em}|=|\mb{\beta}_{ph}c|$ holds for  $\mathbf{S}_{pseu}\ne 0$.  If $\mathbf{S}/W_{em}$ were defined as the group velocity or energy velocity as is done in the classical textbooks \cite{r16,r30}, then the group velocity or energy velocity would be greater than the phase velocity, which is not physical for a non-dispersive lossless medium. 

It is seen from the preceding analysis that the Poynting vector does not necessarily denote a real EM power flow; however, such a phenomenon seems to be neglected in the physics community, in view of the fact that the Abraham momentum, defined through the Poynting vector, is taken as an EM momentum postulate, as proposed by Mansuripur and Zakharian \cite{r18}. 

In summary, we can make some conclusions for the EM momentums, Poynting vector, and EM power flow.
\begin{itemize}
\item  Observed in any inertial frames, the Minkowski EM momentum $\mathbf{D}\times\mathbf{B}$ is parallel to the wave vector (see Eq.\ (\ref{eq37})), which is completely in agreement with the light-momentum criterion as stated in Sec.\ I.  
\item  Observed in the \emph{medium-rest} frame, the Abraham EM momentum $\mathbf{E}\times\mathbf{H}/c^2$  is parallel to the wave vector; however, observed in \emph{general} inertial frames, it is not (see Eq.\ (\ref{eq38})).  Thus the Abraham EM momentum does not meet the light-momentum criterion.
\item  When a medium moves, the Poynting vector $\mathbf{E}\times\mathbf{H}$ consists of two parts: one is parallel to the wave vector, and is a real power flow; the other is perpendicular to the wave vector, and is a pseudo-power flow (see Eq.\ (\ref{eq38})-Eq.\ (\ref{eq40})).
\item Frequency $\omega$, EM energy density $W_{em}=0.5(\mathbf{D}\cdot\mathbf{E}+\mathbf{B}\cdot\mathbf{H})=\mathbf{D}\cdot\mathbf{E}$, phase velocity $\mb{\beta}_{ph}c$, and Poynting vector $\mathbf{E}\times\mathbf{H}$ are related through $\omega (\mathbf{D}\cdot\mathbf{E})>0$ and $(\mb{\beta}_{ph}c)\cdot(\mathbf{E}\times\mathbf{H})=(c/n_d)^2 (\mathbf{D}\cdot\mathbf{E})$, which hold in all inertial frames.
\end{itemize}

\subsection{Four-vector covariance of Minkowski EM momentum and energy, and invariance \\ of the Planck constant}
We have shown the Lorentz covariance of Minkowski photon momentum and energy from the wave four-vector combined with the Einstein light-quantum hypothesis in Sec.\ III.  This covariance suggests that there should be a covariant EM momentum--energy four-vector, given by \\
\begin{equation}
\bar{P}^{\mu}=(\bar{\mathbf{p}}_{em},\bar{E}_{em}/c),
\label{eq42}
\end{equation}  \\
where $\bar{\mathbf{p}}_{em}$  and $\bar{E}_{em}$ are, respectively, the EM momentum and energy for a single ``EM-field cell'' or ``photon'', given by  \\
\begin{equation}
\bar{\mathbf{p}}_{em}=\frac{\mathbf{D}\times\mathbf{B}}{N_p}, \hspace{4mm} \bar{E}_{em}=\frac{\mathbf{D}\cdot\mathbf{E}}{N_p},
\label{eq43}
\end{equation}  \\
with $N_p$ the ``EM-field-cell number density'' or ``photon number density'' in volume.  With $\mathbf{D}\times\mathbf{B}=(\mathbf{D}\cdot\mathbf{E}/\omega)n_d\mathbf{k}$ from Eq.\ (\ref{eq37}) taken into account, the EM momentum--energy four-vector and wave four-vector are related through $\bar{P}^{\mu}=(\bar{E}_{em}/\omega)K^{\mu}$, with  $(\bar{E}_{em}/\omega)=(\mathbf{D}\cdot\mathbf{E})/(N_p\omega)$ corresponding to the Planck constant $\hbar$ physically.  Thus we need to find out the condition for $N_p$  to satisfy for the four-vector covariance of $(\bar{\mathbf{p}}_{em},\bar{E}_{em}/c)$.

The four-vector $\bar{P}^{\mu}=(\bar{\mathbf{p}}_{em},\bar{E}_{em}/c)$  is required to fulfill the four-vector Lorentz rule given by Eqs.\ (\ref{eq1}) and (\ref{eq2}), while the EM fields must fulfill the Lorentz rule of second-rank tensors $F^{\alpha\beta}(\mathbf{E},\mathbf{B})$ and  $G^{\alpha\beta}(\mathbf{D},\mathbf{H})$, given by Eqs.\ (\ref{eq3}) and (\ref{eq4}), or Eqs.\ (\ref{eq22}) and (\ref{eq23}) for a plane wave.  From the four-vector Lorentz transformation of $\bar{P}^{\mu}=(\bar{E}_{em}/\omega)K^{\mu}$, we have \\
\begin{equation}
\frac{N_p\omega}{N'_p\omega'}=\frac{\mathbf{D}\cdot\mathbf{E}}{\mathbf{D}'\cdot\mathbf{E}'} \hspace{3mm} \left(=\frac{W_{em}}{W'_{em}}\right),
\label{eq44}
\end{equation}  \\
which has a clear physical explanation that the Doppler factor of EM energy density is equal to the product of the Doppler factors of EM-field-cell density and frequency.

From the second-rank tensor Lorentz transformations given by Eqs.\ (\ref{eq22}) and (\ref{eq23}), we obtain the transformation of EM energy density $W_{em}=\mathbf{D}\cdot\mathbf{E}~(=\mathbf{B}\cdot\mathbf{H})$, given by \\
\begin{equation}
\frac{(\mathbf{D}\cdot\mathbf{E})}{(\mathbf{D}'\cdot\mathbf{E}')}=\gamma(1-n'_d\mathbf{\hat{n}}'\cdot\mb{\beta}')
\left[\gamma\left(1-\frac{1}{n'_d}\mathbf{\hat{n}}'\cdot\mb{\beta}'\right)\right],
\label{eq45}
\end{equation}  \\
namely, Eq.\ (\ref{eq33}).  Comparing with Eq.\ (\ref{eq6}), we know that $\gamma(1-n'_d\mathbf{\hat{n}}'\cdot\mb{\beta}')$  is the frequency Doppler factor.  In free space ($n'_d=1$), Eq.\ (\ref{eq45}) is reduced to Einstein's result \cite{r20}:  $|\mathbf{E}|=\gamma(1-\mathbf{\hat{n}}'\cdot\mb{\beta}')|\mathbf{E}'|$.

Inserting Eq.\ (\ref{eq45}) into Eq.\ (\ref{eq44}) we obtain the transformation of the EM-field-cell density $N_p$, given by \\
\begin{equation}
N_p=\gamma\left(1-\frac{1}{n'_d}\mathbf{\hat{n}}'\cdot\mb{\beta}'\right)N'_p.
\label{eq46}
\end{equation}  \\
\indent So far we have finished the proof of the covariance of $(\bar{\mathbf{p}}_{em},\bar{E}_{em}/c)$ by resorting to a parameter of $N_p$, so-called “EM-field-cell density”.  Actually, we do not have to know what the specific value of $N'_p$ or $N_p$ is, but the ratio of $N_p/N'_p$, and $\bar{P}^{\mu}=(\bar{\mathbf{p}}_{em},\bar{E}_{em}/c)$ is pure ``classical'', without Planck constant $\hbar$ involved.  However, $N_p$ must be the ``photon density'' when the Einstein light-quantum hypothesis is imposed.  In such a case, we have $\hbar=(\mathbf{D}\cdot\mathbf{E})/(N_p\omega)$, or $N_p=[(\mathbf{D}_0\cdot\mathbf{E}_0)/(\hbar\omega)]\cos^2\Psi$, namely, the photon density $N_p$  is a ``wave''.

Mathematically speaking, the existence of the covariance of $\bar{P}^{\mu}=(\bar{\mathbf{p}}_{em},\bar{E}_{em}/c)$ is apparent.   $\bar{P}^{\mu}$ and $K^{\mu}$  are ``parallel'', and differ only by a factor of $(\bar{E}_{em}/\omega)=(\mathbf{D}\cdot\mathbf{E})/(N_p\omega)$, which contains an introduced parameter $N_p$ to make the transformation hold.  

It is seen from Eqs.\ (\ref{eq43}) and (\ref{eq44}) that $(\bar{E}_{em}/\omega)=(\bar{E}'_{em}/\omega')$ holds and is a Lorentz invariant, and $(\bar{E}_{em}/\omega)=\hbar$ holds when the Einstein light-quantum hypothesis is imposed.  Thus the Planck constant $\hbar$  must be Lorentz invariant.  In other words, the Einstein light-quantum hypothesis requires the Lorentz invariance of the Planck constant for a plane wave.  Therefore, the construction of the photon momentum--energy four-vector, Eq.\ (\ref{eq16}) in Sec.\ III, is well grounded.

If a volume $dV'_{\mathrm{light}}$  in the medium-rest frame moves along the wave vector $n'_d\mathbf{k}'$  at light speed $(c/n'_d)$, then there are no photons that cross its boundary, and the photon number within $dV'_\mathrm{light}$ remains constant.  In such a case, the transformation of the moving volume (termed \emph{light volume}) is given by \\
\begin{equation}
\frac{dV'_\mathrm{light}}{dV_\mathrm{light}}=\gamma\left(1-\frac{1}{n'_d}\mathbf{\hat{n}}'\cdot\mb{\beta}'\right).
\label{eq47}
\end{equation}  \\
Comparing with Eq.\ (\ref{eq46}), we find that \\
\begin{equation}
N_p dV_\mathrm{light}=N'_p dV'_\mathrm{light}
\label{eq48}
\end{equation}  \\
is Lorentz invariant, namely, the photon number in the light volume is Lorentz invariant.  Thus we have the total momentum--energy four-vector in the light volume, given by \\
\begin{equation}
\bar{P}^{\mu}(N_p dV_\mathrm{light})=(\mathbf{D}\times\mathbf{B},\mathbf{D}\cdot\mathbf{E}/c)dV_\mathrm{light},
\label{eq49}
\end{equation}  \\
or
\begin{equation}
\int\limits_{V_\mathrm{light}}(\bar{P}^{\mu}N_p)dV=\int\limits_{V_\mathrm{light}}(\mathbf{D}\times\mathbf{B},\mathbf{D}\cdot\mathbf{E}/c)dV.
\label{eq50}
\end{equation}  \\
\indent The invariance of $N_p dV_\mathrm{light}$  implies that, observed in any inertial frames, all the $N_p dV_\mathrm{light}$  photons are frozen inside the light volume  $dV_\mathrm{light}$.  Thus the light volume can be taken to be an approximate description of practical low-divergence light pulses. 

Inserting Eq.\ (\ref{eq47}) and Eq.\ (\ref{eq6}) into Eq.\ (\ref{eq45}), we have \\
\begin{equation}
\frac{(\mathbf{D}\cdot\mathbf{E})dV_\mathrm{light}}{\omega}=\frac{(\mathbf{D}'\cdot\mathbf{E}')dV'_\mathrm{light}}{\omega'},
\label{eq51}
\end{equation}  \\
which is also a Lorentz invariant; namely, the light-volume energy and the frequency transform in the same law.  This result, which is obtained in the moving medium, is exactly the same as that obtained by Einstein in \emph{free-space} \cite{r20}. 

From the preceding analysis, we can draw the following conclusions.
\begin{itemize}

\item The Minkowski momentum per unit EM-field cell,  $N_p^{-1}(\mathbf{D}\times\mathbf{B})$, is Lorentz covariant, as the space component of the EM momentum--energy four-vector $\bar{P}^{\mu}=N_p^{-1}(\mathbf{D}\times\mathbf{B},\mathbf{D}\cdot\mathbf{E}/c)$, just like the Minkowski photon momentum $\hbar n_d\mathbf{k}$ is Lorentz covariant, as the space component of the photon momentum--energy four-vector $P^{\mu}=(\hbar n_d\mathbf{k},\hbar\omega/c)$.  When Einstein light-quantum hypothesis $N_p^{-1}\mathbf{D}\cdot\mathbf{E}=\hbar\omega$  is imposed on the former, the two four-vectors become the same, namely,  $N_p^{-1}(\mathbf{D}\times\mathbf{B},\mathbf{D}\cdot\mathbf{E}/c)=(\hbar n_d\mathbf{k},\hbar\omega/c)$.

\item There are two forms of momentum--energy four-vectors: (a) the momentum and energy in a single EM-field cell or photon constitute a four-vector, namely, $N_p^{-1}(\mathbf{D}\times\mathbf{B},\mathbf{D}\cdot\mathbf{E}/c)$  or $(\hbar n_d\mathbf{k},\hbar\omega/c)$ is a four-vector; and (b) the total momentum and energy in a light volume constitute a four-vector, namely, $(\mathbf{D}\times\mathbf{B},\mathbf{D}\cdot\mathbf{E}/c)dV_\mathrm{light}$  is also a four-vector.  However, the momentum and energy densities themselves cannot directly constitute a four-vector, namely, $(\mathbf{D}\times\mathbf{B},\mathbf{D}\cdot\mathbf{E}/c)$ or $N_p(\hbar n_d\mathbf{k},\hbar\omega/c)$ is never a Lorentz four-vector.

\item The Planck constant is a Lorentz invariant, which is a strict result of special relativity and quantized light energy for a plane wave in a moving uniform medium.  Obviously, this conclusion is also valid in free space because the empty space is a special kind of ``uniform medium''.
\end{itemize}

\subsection{Issue of light momentum transfer and indeterminacy of EM momentum \\conservation equation}
Now let us consider the issue of EM momentum transfer between a plane wave and medium, and find out why the momentum conservation equation cannot be uniquely determined in the Maxwell-equation frame without resorting to the principle of relativity. 

From Eqs.\ (\ref{eq24}) and (\ref{eq25}), with $\mathbf{J}=0$ and $\rho=0$ taken into account we have \\
\begin{equation}
\frac{\partial(\mathbf{D}\times\mathbf{B})}{\partial t}=-\mathbf{D}\times(\nabla\times\mathbf{E})-\mathbf{B}\times(\nabla\times\mathbf{H}).
\label{eq52}
\end{equation} \\
Because~ $(\mathbf{E},~\mathbf{B},~\mathbf{D},~\mathbf{H})=(\mathbf{E}_0,~\mathbf{B}_0,~\mathbf{D}_0,~\mathbf{H}_0)\cos\Psi$, with~ $\Psi=(\omega t-n_d\mathbf{k}\cdot\mathbf{x})$,~ $\nabla\cdot\mathbf{D}=0\Rightarrow\mathbf{D}\bot\mathbf{\hat{n}}$, $\nabla\cdot\mathbf{B}=0\Rightarrow\mathbf{B}\bot\mathbf{\hat{n}}$, and $\mathbf{D}\cdot\mathbf{E}=\mathbf{B}\cdot\mathbf{H}$, leading to $\partial(\mathbf{D}\times\mathbf{B})/\partial t=-2(\mathbf{D}_0\cdot\mathbf{E}_0)(\cos\Psi\sin\Psi)(n_d\mathbf{k})$, we have \\
\begin{equation}
\frac{\partial(\mathbf{D}\times\mathbf{B})}{\partial t}=-\nabla\cdot\check{\mathbf{T}}_M,
\label{eq53}
\end{equation} \\
where the symmetric Minkowski EM stress tensor is given by \\
\begin{equation}
\mathbf{\check{T}}_M=\mathbf{\check{I}}(\mathbf{D}\cdot\mathbf{E})=\mathbf{\check{I}}(\mathbf{B}\cdot\mathbf{H}),
\label{eq54}
\end{equation} \\ 
where $\mathbf{\check{I}}$ is the unit tensor.

Equation\ (\ref{eq53}) is the Minkowski momentum conservation equation for a plane wave in a moving medium, and is invariant in form in all inertial frames together with the Maxwell equations.

To understand the physical implication of stress tensor, let us consider the EM momentum in a given dielectric volume $V$ closed by surface $S$.  From Eq.\ (\ref{eq53}), using divergence theorems for a tensor and a vector, with the stress tensor $\mathbf{\check{T}}_M=\mathbf{\check{I}}(\mathbf{D}_0\cdot\mathbf{E}_0)\cos^2 (\omega t-n_d\mathbf{k}\cdot\mathbf{x})$ inserted, we have \\
\begin{align}
\frac{\partial}{\partial t}\int_V&(\mathbf{D}\times\mathbf{B})dV=-\oint_S d\mathbf{S}\cdot\check{\mathbf{T}}_M  \nonumber
\\
&=-\oint_S d\mathbf{S}(\mathbf{D}\cdot\mathbf{E}) \nonumber=-\int_V \nabla(\mathbf{D}\cdot\mathbf{E})dV \nonumber
\\
&=-\int_V (\mathbf{D}_0\cdot\mathbf{E}_0)\sin(2\Psi)(n_d\mathbf{k})dV,
\label{eq55}
\end{align}  \\
where $\int(\mathbf{D}\times\mathbf{B})dV$ is the total EM momentum in $V$, and $d\mathbf{S}\cdot\check{\mathbf{T}}_M$ is the momentum element through the $d\mathbf{S}$-area element per unit time.  From Eq.\ (\ref{eq55}), we find  $\oint d\mathbf{S}\cdot\check{\mathbf{T}}_M\ne 0$ in general while the time average $<\oint d\mathbf{S}\cdot\check{\mathbf{T}}_M>=0$, which also hold in \emph{empty space}.  This phenomenon results from the ``travelling-wave'' attribution of $\check{\mathbf{T}}_M$; namely, $\check{\mathbf{T}}_M$ varies with space coordinates, and the total EM momentums flowing in ($d\mathbf{S}\cdot\check{\mathbf{T}}_M<0$ ) and out ($d\mathbf{S}\cdot\check{\mathbf{T}}_M>0$) of a given volume $V$ are usually different at a given instant, but they are equal on the time average.  In other words, $\partial/\partial t\int(\mathbf{D}\times\mathbf{B})dV$ only denotes the change rate of momentum flowing into $V$ at a given instant, instead of a force given by the medium; thus \emph{there is no momentum transfer taking place between the plane wave and the uniform medium}, and there is no force acting on the dielectric.  This also can be understood through the light-quantized Minkowski EM-field-cell or photon four-vector $N_p^{-1}(\mathbf{D}\times\mathbf{B},\mathbf{D}\cdot\mathbf{E}/c)=(\hbar n_d\mathbf{k},\hbar\omega/c)$, which indicates that the momentum of a photon remains constant during propagation in a uniform medium.  Obviously, this conclusion is applicable to any inertial frames and can be used to explain why the momentum transfer only takes place on the vacuum-medium interface in the medium Einstein-box thought experiment for a light pulse \cite{r21}.

It should be pointed out that the conventional EM force definition (see p. 159 of Ref. \cite{r27}, for example) is questionable because it cannot pass a plane-wave test.  For a plane wave in the medium-rest frame $X'Y'Z'$, the medium is isotropic and uniform ($\partial\epsilon'/\partial\mathbf{x}'=0$ and $\partial\mu'/\partial\mathbf{x}'=0$).  According to the conventional definition, however, the EM force exerted on a volume is given by $\mathbf{f}'=[(n'^2_d-1)/c^2]\partial(\mathbf{E}'\times\mathbf{H}')/\partial t'$, which implies that there is a momentum transfer between the plane wave and the uniform medium, clearly contradicting the conclusion obtained from above.  Thus this conventional definition \cite{r27} is flawed.

The construction of the stress tensor is flexible; in a sense, it is artificial within the Maxwell-equation frame.  For example, $\nabla\cdot(-\mathbf{DE}-\mathbf{BH})=0$ and $\mathbf{D}\cdot\mathbf{E}=\mathbf{B}\cdot\mathbf{H}$  hold for a plane wave, and the Minkowski tensor can be rewritten in an asymmetric form \cite{r14} \\
\begin{equation}
\check{\mathbf{T}}_M=-\mathbf{DE}-\mathbf{BH}+\check{\mathbf{I}}\frac{1}{2}(\mathbf{D}\cdot\mathbf{E}+\mathbf{B}\cdot\mathbf{H}),
\label{eq56}
\end{equation} \\
which does not affect the validity of Eq.\ (\ref{eq53}) and Eq.\ (\ref{eq55}).

Similarly, we can obtain the Abraham momentum conservation equation, given by \\
\begin{equation}
\frac{\partial}{\partial t}\left(\frac{\mathbf{E}\times\mathbf{H}}{c^2}\right)=-\nabla\cdot\mathbf{\check{T}}_A,
\label{eq57}
\end{equation} \\
where the Abraham stress tensor is given by \\
\begin{equation}
\check{\mathbf{T}}_A=\beta^2_{ph}[-(\mathbf{ED}+\mathbf{HB})+\check{\mathbf{I}}(\mathbf{D}\cdot\mathbf{E})],
\label{eq58}
\end{equation} \\
which is not symmetric.  By taking advantage of $\nabla\cdot(\mathbf{DE})=0$,  $\nabla\cdot(\mathbf{BH})=0$, and $\mathbf{D}\cdot\mathbf{E}=\mathbf{B}\cdot\mathbf{H}$, the Abraham stress tensor can be rewritten in a symmetric form, given by

\begin{align}
\check{\mathbf{T}}_A=&\beta^2_{ph} [ -(\mathbf{ED}+\mathbf{DE})  \nonumber 
\\
&-(\mathbf{HB}+\mathbf{BH})~+~\check{\mathbf{I}}\frac{1}{2}(\mathbf{D}\cdot\mathbf{E}+\mathbf{B}\cdot\mathbf{H})~].
\label{eq59}
\end{align}

Note that in a moving medium, $\mathbf{ED}=\mathbf{DE}$ and $\mathbf{HB}=\mathbf{BH}$  usually are not true because of the anisotropy of the moving medium.  However, in free space, $\mathbf{ED}=\mathbf{DE}$ and $\mathbf{HB}=\mathbf{BH}$ always hold because the empty space ($\beta^2_{ph}=1$) is isotropic observed in any inertial frames.  Thus in free space, $\check{\mathbf{T}}_M$  given by Eq.\ (\ref{eq56}) and $\check{\mathbf{T}}_A$  given by Eq.\ (\ref{eq58}) are identical.  

It can be seen from Eq.\ (\ref{eq53}) and Eq.\ (\ref{eq57}) that the momentum conservation equations are all differential equations and they can be converted one to the other through the Maxwell equations.  In fact, Eq.\ (\ref{eq53}) and Eq.\ (\ref{eq57}) can be obtained from a more general momentum conservation equation given by $\partial\mathbf{g}/\partial t+\nabla\cdot\check{\mathbf{T}}=0$, where $\mathbf{g}=a\mathbf{g}_A+(1-a)\mathbf{g}_M$ and  $\check{\mathbf{T}}=a\check{\mathbf{T}}_A+(1-a)\check{\mathbf{T}}_M$, with $\mathbf{g}_A=(\mathbf{E}\times\mathbf{H})/c^2$,  $\mathbf{g}_M=\mathbf{D}\times\mathbf{B}$,  $a$ being any constant, $\check{\mathbf{T}}_M$ being given by Eq.\ (\ref{eq56}), and $\check{\mathbf{T}}_A$ being given by Eq.\ (\ref{eq58}).  We have  $\partial\mathbf{g}_M/\partial t+\nabla\cdot\check{\mathbf{T}}_M=0$, namely, Eq.\ (\ref{eq53}) for $a=0$, and  $\partial\mathbf{g}_A/\partial t+\nabla\cdot\check{\mathbf{T}}_A=0$, namely, Eq.\ (\ref{eq57}) for  $a=1$, while $\mathbf{g}$ and $\check{\mathbf{T}}$ are restored to $\mathbf{g}=\mathbf{g}_A=\mathbf{g}_M$ and $\check{\mathbf{T}}=\check{\mathbf{T}}_A=\check{\mathbf{T}}_M$ in free space.  Just as indicated in Sec.\ I, the Maxwell equations themselves support various forms of momentum conservation equations, resulting in an indeterminacy of momentum definitions.  Thus, to identify the correctness of momentum definitions, Fermat's principle and the principle of relativity are indispensable.

\subsection{Negative frequency and negative EM energy density for a superluminal medium}
In addition to the negative-frequency appearance, as indicated by Huang \cite{r28}, a negative energy density (NED) may result for a plane wave when the medium moves opposite to the wave vector direction at a faster-than-dielectric light speed; this phenomenon is called ``NED zone'' for the sake of convenience.

In the NED zone, the photons possess negative energy from the viewpoint of phenomenological quantum-electrodynamics \cite{r31}.  The origin of negative EM energy density can be seen from Eq.\ (\ref{eq45}).  The energy density Doppler factor is equal to the product of the Doppler factors of frequency and EM-field-cell density.  The EM-field-cell density Doppler factor is always positive while the frequency Doppler factor is negative in the NED zone ($n'_d\mathbf{\hat{n}}'\cdot\mb{\beta}'=n'_d|\mb{\beta}'|>1$), leading to $\mathbf{D}\cdot\mathbf{E}<0$  when $\omega<0$.  (Note that no matter whether $\omega<0$  or $\omega>0$, Eqs.\ (\ref{eq37}) and (\ref{eq39}) can be written as $\mathbf{D}\times\mathbf{B}=\mathbf{\hat{n}}|(\mathbf{D}\cdot\mathbf{E})|(n_d/c)$ and $\mathbf{S}_{power}=\mathbf{\hat{n}}|(\mathbf{D}\cdot\mathbf{E})|(c/n_d)$, with $\mathbf{S}_{power}=(c/n_d)^2(\mathbf{D}\times\mathbf{B})$.) 

In the NED zone,  $\mathbf{D}\|\mathbf{E}$ and $\mathbf{H}\|\mathbf{B}$ are valid, and from Eqs.\ (\ref{eq22}) and (\ref{eq23}) we have  \\
\begin{align}
&\frac{\mathbf{D}}{\mathbf{E}}=\left(\frac{1-|\mb{\beta}'|/n'_d}{1-n'_d|\mb{\beta}'|}\right)\epsilon'<0,
\label{eq60}
\\  \nonumber
\\  
&\frac{\mathbf{B}}{\mathbf{H}}=\left(\frac{1-|\mb{\beta}'|/n'_d}{1-n'_d|\mb{\beta}'|}\right)\mu'<0.
\label{eq61}
\end{align}  \\
\indent Because of $\omega<0$  in the NED zone, from Eq.\ (\ref{eq11}) we have $\mb{\beta}_{ph}c=(\beta_{ph}c)\mathbf{\hat{n}}$ with $\beta_{ph}c<0$.  From Eqs.\ (\ref{eq35}) and (\ref{eq36}) we have  \\
\begin{align}
&\mathbf{E}=-\beta_{ph}c~\mathbf{\hat{n}}\times\mathbf{B},
\label{eq62}
\\  
&\mathbf{H}=+\beta_{ph}c~\mathbf{\hat{n}}\times\mathbf{D}.
\label{eq63}
\end{align}  \\
\indent It follows that the plane wave in the NED zone is a left-hand wave: (a) $(\mathbf{E}, \mathbf{B}, \mathbf{\hat{n}})$ and $(\mathbf{D}, \mathbf{H}, \mathbf{\hat{n}})$  follow the left-hand rule, and (b) the phase velocity or group velocity is opposite to the wave vector.  In other words, the moving medium in the NED zone behaves as a so-called ``negative index medium'' \cite{r32}, where the refractive index is taken to be negative, instead of the frequency here.  \\
\indent As mentioned above, in the NED zone the group velocity $\mathbf{v}_{gr}$ is opposite to the wave vector $n_d\mathbf{k}$, and $\mathbf{D}\cdot\mathbf{E}<0$ holds.  Thus from Eq.\,(\ref{eq38})--Eq.\,(\ref{eq40}), with $\mathbf{S}_{pseu}=0$ taken into account, we find that the Poynting vector $\mathbf{S}=\mathbf{S}_{power}=(\mathbf{D}\cdot\mathbf{E})\mathbf{v}_{gr}$ also has the same direction as the wave vector $n_d\mathbf{k}$ or the EM momentum $\mathbf{D}\times\mathbf{B}$ for the ``negative index medium'' effect.  This conclusion makes sense physically, because photons are the carriers of EM energy and momentum, and the EM power flow and momentum are supposed to have the same direction. \\
\indent As we have known, the NED zone results from  $\omega<0$.  However, understanding the sign of the frequency has been thought to be a difficult question.  Just as Huang indicated \cite{r28}, ``Can we find a convincing explanation of the meaning of negative frequency of waves in any literature?''  In fact, from the following analysis we can see that the sign of the frequency (EM energy density) is just a reflection of the property of wave propagation. \\
\indent In the NED zone we have $n_d\mathbf{k}=(1-n'^{-1}_d\mb{\beta}'\cdot\mathbf{\hat{n}}')\gamma(n'_d\mathbf{k}')$.  Because of $(1-n'^{-1}_d\mb{\beta}'\cdot\mathbf{\hat{n}}')>0$, $(n_d\mathbf{k})$  and $(n'_d\mathbf{k}')$  take the same direction.  Observed in the laboratory frame and the medium-rest frame, respectively, both power flows have the same direction.  From Eqs.\ (\ref{eq60}) and (\ref{eq61}) we see that the moving medium in such a case physically behaves as a ``negative index medium''.  Thus the negative-frequency effect denotes a distinct physical phenomenon where the EM wave is a left-hand wave.  In other words, the sign of the frequency (EM energy density) only characterizes the propagation property of EM waves.  Experimentally, the observed frequency is always positive, and a positive EM energy propagates along the wave vector direction, while the sign of the frequency is determined by examining the property of wave propagation in the moving medium: (--) for the left-hand and (+) for the right-hand. \\
\indent It is interesting to point out that, in the effect of the ``negative index medium'' analyzed herein, dispersion of the medium material is not required, and the Poynting vector, EM momentum, and wave vector all have the \emph{same} direction.  In contrast, in the traditional effect of negative index medium, which was first analyzed by Veselago, the medium material must be dispersive to support a positive EM energy, and the Poynting vector is directed \emph{opposite} to the EM momentum or wave vector \cite{r32}. \\
\indent In summary, we have shown that when the medium moves opposite to the wave vector at a faster-than-dielectric light speed (superluminal medium), two interrelated phenomena occur: (a) the frequency and EM energy density become negative; and (b) the plane wave is left-handed, with the phase velocity opposite to the Poynting vector, which is called a ``negative phase velocity'' according to the definition in Ref.\ \cite{r33}. \\
\indent The appearance of the negative phase velocity in the superluminal medium can be directly understood from the expressions:  $\omega (\mathbf{D}\cdot\mathbf{E})>0$ and $(\mb{\beta}_{ph}c)\cdot(\mathbf{E}\times\mathbf{H})=(c/n_d)^2(\mathbf{D}\cdot\mathbf{E})$, which hold in all inertial frames, as shown in Sec.\ IV\,A. In the medium-rest frame, the plane wave is a right-hand wave with frequency $\omega'>0$, EM energy density $\mathbf{D}'\cdot\mathbf{E}'=\epsilon' |\mathbf{E}'|^2>0$, and $\mathbf{E}'\times\mathbf{H}'=(\mathbf{D}'\cdot\mathbf{E}')(\mb{\beta}'_{ph}c)$ $\Rightarrow$ the phase velocity $\mb{\beta}'_{ph}c$  having the same direction as the Poynting vector $\mathbf{E}'\times\mathbf{H}'$  (positive phase velocity).  However, in the superluminal medium, with $\omega<0\Rightarrow(\mathbf{D}\cdot\mathbf{E})<0$  we have $(\mb{\beta}_{ph}c)\cdot(\mathbf{E}\times\mathbf{H})=(c/n_d)^2(\mathbf{D}\cdot\mathbf{E})<0$, namely $\mb{\beta}_{ph}c$  is a negative phase velocity.  \\
\indent It should be pointed out that the negative phase velocity and negative refraction in a uniformly moving medium have been well studied in a number of recent publications \cite{r33,r34,r35}.  Mackay and Lakhtakia concluded that a plane wave with a \emph{positive} phase velocity, observed in the medium-rest frame, may become a plane wave with \emph{negative} phase velocity observed in moving frames \cite{r33}. Mackay--Lakhtakia's conclusion is clearly supported by the example given in the present paper, as indicated above.  However, it should be noted that in all those analyses \cite{r33,r34,r35}, the frequency is always taken to be positive, and thus the negative frequency and negative EM energy density appearing in the superluminal medium, presented in the present paper, have never been clearly exposed, and their physics has never been formulated.    

\section{Conclusions and Remarks}
An EM plane wave, although not practical, is the simplest strict solution of Maxwell equations, and it is often used to explore fundamental physics.  For example, Einstein used a plane wave to develop the special theory of relativity and derived the well-known relativistic Doppler formula in free space \cite{r20}.  In this paper, we use the plane wave, which propagates in a moving non-dispersive, lossless, non-conducting, isotropic, uniform medium, to determine which formulation of light momentum is correct.  

We have shown that (a) Minkowski light momentum and energy constitute a Lorentz four-vector, while Abraham momentum and energy do not; and (b) observed in any inertial frame, Minkowski EM momentum $\mathbf{g}_M=\mathbf{D}\times\mathbf{B}$ always take the direction of the wave vector $n_d\mathbf{k}$, while the Abraham momentum $\mathbf{g}_A=\mathbf{E}\times\mathbf{H}/c^2$  does not, unless in free space or when the dielectric medium moves parallel to the wave vector, as shown in Eqs.\ (\ref{eq37}) and (\ref{eq38}).  The Minkowski momentum is completely consistent with Fermat's principle and the principle of relativity, and it is the unique correct light momentum.

The photon momentum--energy four-vector $P^{\mu}=\hbar K^{\mu}$ is constructed based on the wave four-vector combined with Einstein light-quantum hypothesis, while the EM momentum--energy four-vector $\bar{P}^{\mu}=N^{-1}_p(\mathbf{D}\times\mathbf{B}, \mathbf{D}\cdot\mathbf{E}/c)$  is constructed based on the Lorentz covariance of EM field-strength tensors $F^{\alpha\beta}(\mathbf{E},\mathbf{B})$  and $G^{\alpha\beta}(\mathbf{D},\mathbf{H})$ (to keep Maxwell equations invariant in form in all inertial frames) \cite{r25}, where $N^{-1}_p\mathbf{D}\times\mathbf{B}$ and $N^{-1}_p\mathbf{D}\cdot\mathbf{E}$ are, respectively, the momentum and energy for a ``single EM-field cell''.  When the Einstein light-quantum hypothesis  $N^{-1}_p\mathbf{D}\cdot\mathbf{E}=\hbar\omega$  is imposed on the latter, the latter is restored to the former, namely, $N^{-1}_p(\mathbf{D}\times\mathbf{B}, \mathbf{D}\cdot\mathbf{E}/c)=(\hbar n_d\mathbf{k}, \hbar\omega/c)$, with the ``single EM-field cell'' becoming ``single photon''.  Thus the single photon momentum is the direct result of Einstein light-quantized EM momentum. In other words, the monochromatic plane wave is an identical-photon model, in which every photon has the same momentum and energy so that the light-wave momentum has the exact and simplest definition. \\
\indent As the carriers of EM momentum and energy of the plane wave, all photons move uniformly in any inertial frames at the phase velocity $\mb{\beta}_{ph}c=(\omega/|n_d\mathbf{k}|)\mathbf{\hat{n}}$ (= group velocity = energy velocity).  There is no momentum transfer taking place between the plane wave and the uniform medium, and there is no EM force acting on the medium.  The photon density $N_p=W_{em}/(\hbar\omega)$  is a ``particle density wave'', and the EM power flow  $\mathbf{S}_{power}=W_{em}\mb{\beta}_{ph}c=\mathbf{\hat{n}}|W_{em}|(c/n_d)=\mathbf{\hat{n}}|N_p\hbar\omega|(c/n_d)$  is also a ``wave'', changing with time and space, where $W_{em}=\mathbf{D}\cdot\mathbf{E}=\mathbf{B}\cdot\mathbf{H}=(\mathbf{D}_0\cdot\mathbf{E}_0)\cos^2 (\omega t-n_d\mathbf{k}\cdot\mathbf{x})$  is the EM energy density.   \\
\indent The principle of relativity, Fermat's principle, and the global momentum--energy conservation law are all basic postulates in physics.  In the principle-of-relativity frame, it is Fermat's principle that requires the correct light momentum and energy to constitute a Lorentz four-vector for a plane wave in a moving uniform medium, as shown in this paper, while it is the global momentum--energy conservation law that requires the correct light momentum and energy to constitute a Lorentz four-vector in a medium Einstein-box thought experiment as shown in Ref.\ \cite{r21}.  From this we can conclude that the justification of Minkowski momentum as the correct light momentum is completely required by the basic postulates in physics. \\
\indent It should be emphasized that the significance of the resolution of the Abraham--Minkowski debate presented in the paper is not just to show the justification of the Minkowski momentum as the unique correct light momentum.  In fact, through seeking the resolution we have clarified and developed some basic concepts and principles in electrodynamics and special relativity, which are outlined as follows. \vspace{3mm} \\
\indent \textbf{(i) Light-momentum criterion.}  We have set up a light-momentum criterion for the first time, which states that the momentum of light in a medium (including empty space) is parallel to the wave vector in all inertial frames.  This criterion is the direct result of the principle of relativity and Fermat's principle for a plane wave.  In a dielectric medium (not including empty space), Minkowski momentum satisfies the criterion while Abraham momentum does not; thus the Minkowski momentum is the unique correct light momentum.  In empty space, Minkowski and Abraham momentums are equal, and both satisfy the criterion.  \\
\indent It is worthwhile to point out that this light-momentum criterion provides a necessary \emph{physical} condition to find out whether a \emph{mathematical} expression can represent the correct momentum of light.  This is illustrated as follows.  \\
\indent Conventionally, the EM momentum--energy stress tensor is used to define the EM momentum of light in a medium. Minkowski first developed an EM tensor, corresponding to Minkowski momentum $\mathbf{D}\times\mathbf{B}$, and later, Abraham also suggested an EM tensor, corresponding to Abraham momentum $\mathbf{E}\times\mathbf{H}/c^2$ \cite{r13}.  It is generally argued that Maxwell equations are manifestly Lorentz covariant while the EM tensor follows from the Maxwell equations; thus the EM momentum defined from the EM tensor certainly respects the principle of relativity.  However, it should be indicated that such an argument is based on an incomplete understanding of the relativity principle.  The reason is as follows.  \\
\indent The relativity principle states that physical laws are the same in all inertial frames of reference.  A physical law has its specific physical implication, which is usually expressed through mathematical equations.  Thus the relativity principle requires that the mathematical equations describing the law must be the same in form in all inertial frames, and the specific physical implication of the equations also must be the same.  As far as the \emph{mathematical expression} is concerned, $\mathbf{D}\times\mathbf{B}$ and $\mathbf{E}\times\mathbf{H}/c^2$, which respectively satisfy their own momentum conservation equations, are both Lorentz covariant together with the Maxwell equations (namely, the same mathematical forms in all frames); however, as far as the \emph{physical implication} is concerned, only $\mathbf{D}\times\mathbf{B}$ is Lorentz covariant (the same physical implication in all frames) while $\mathbf{E}\times\mathbf{H}/c^2$ is not (not the same physical implication in all frames).  That is because $\mathbf{D}\times\mathbf{B}$ meets the light-momentum criterion while $\mathbf{E}\times\mathbf{H}/c^2$ does not.  \vspace{2mm} \\
\indent \textbf{(ii) Poynting vector.}  In conventional EM wave theory, the Poynting vector $\mathbf{S}=\mathbf{E}\times\mathbf{H}$ as EM power flow has been thought to be a well-established basic concept.  In view of the existence of some kind of mathematical ambiguity for this concept, some scientists suggested it to be a ``postulate'' \cite{r18}, or ``hypothesis'', ``until a clash with new experimental evidence shall call for its revision'' \cite[p.135]{r27}.  However, in this paper, we have shown that the Poynting vector may not denote the EM power flow in an anisotropic medium (see Eq.\,(\ref{eq38})--Eq.\,(\ref{eq40})).  This result revises the conventional understanding of Poynting vector, and also explains why Laue--M\o ller theory \cite{r15,r16} and Mansuripur--Zakharian theory \cite{r18} have the same Poynting-vector assumption but have completely different physical results: one supporting Minkowski momentum and the other supporting Abraham momentum.    \vspace{2mm} \\
\indent \textbf{(iii) Force exerted by an EM field in a medium.}  It is well accepted conventionally that the force exerted by an EM field on a unit volume of isotropic dielectric medium is given by \cite{r12,r27,r36,r37,r38,r39}  \\
\begin{equation}
\mathbf{f}=\mathbf{f}^M+\mathbf{f}^A,~~~~~~~~~~~~~~~~~~~~~~~~~~~~~~~~~~
\label{eq64}
\end{equation} 
where 
\begin{align}
&\mathbf{f}^M=\rho\mathbf{E}+\mathbf{J}\times\mathbf{B}-\frac{1}{2}\mathbf{E}^2\nabla\epsilon-\frac{1}{2}\mathbf{H}^2\nabla\mu,
\label{eq65}
\\
&\mathbf{f}^A=\frac{n^2_d-1}{c^2}\frac{\partial}{\partial t}(\mathbf{E}\times\mathbf{H}).
\label{eq66}
\end{align}  \\						
It is argued that the $\mathbf{f}^A$-term ``simply fluctuates out when averaged over an optical period in a stationary beam'', but ``it is in principle measurable''\cite{r36}, and various ideas were proposed to experimentally measure or identify the $\mathbf{f}^A$-term \cite{r37,r38}.  Unfortunately, as shown herein, the preceding conventional formula cannot pass the test of a plane wave in an isotropic, lossless, uniform medium, and consequently, it is flawed.  To directly understand this, inserting  $\rho=0$,  $\mathbf{J}=0$,  $\nabla\epsilon=0$,  $\nabla\mu=0$, and $(\mathbf{E}, \mathbf{H})=(\mathbf{E}_0, \mathbf{H}_0)\cos (\omega t-n_d \mathbf{k}\cdot\mathbf{x})$ into Eq.\ (\ref{eq64}), we have $\mathbf{f}^M=0$ and \\
\begin{equation}
\mathbf{f}=\frac{n^2_d-1}{c^2}(\mathbf{E}_0\times\mathbf{H}_0)\frac{\partial}{\partial t}\cos^2 (\omega t-n_d\mathbf{k}\cdot\mathbf{x}) \ne 0
\label{eq67}
\end{equation} \\
holding (except for those discrete points).  According to the physical implication of force, $\mathbf{f}\ne 0$ implies that there is a momentum transfer taking place between the plane wave and the medium.  However, this apparently contradicts the fact that in an isotropic uniform medium ($\nabla\epsilon=0$ and $\nabla\mu=0$), all photons move uniformly and they do not have any momentum exchanges with the medium.  Thus this conventional EM force formula Eq.\ (\ref{eq64}) is indeed flawed.   \vspace{2mm} \\
\indent \textbf{(iv) Total-momentum model.}  We have shown in Ref.\ \cite{r21} that the total-momentum model proposed by Barnett \cite{r19}, which is widely accepted in the community, is not compatible with the principle of relativity and the global momentum--energy conservation law, which are all fundamental postulates in physics.  This conclusion comes from the fact that in the Einstein-box thought experiment, the medium-box kinetic momentum and energy always constitute a Lorentz four-vector before and after the photon enters the box because the medium-box is made up of \emph{massive particles}, while the Abraham photon momentum and energy cannot constitute a Lorentz four-vector after the photon enters the box; resulting in the breakdown of the total momentum--energy conservation law within the principle-of-relativity frame.  \\
\indent In the total-momentum model, as shown by Eq.\ (7) of Ref.\ \cite{r19}, the total momentum is given by $\mathbf{p}^{\mathrm{med}}_{\mathrm{kin}}+\mathbf{p}_{\mathrm{Abr}}
=\mathbf{p}^{\mathrm{med}}_{\mathrm{can}}+\mathbf{p}_{\mathrm{Min}}$, where $\mathbf{p}_{\mathrm{Abr}}$ and $\mathbf{p}_{\mathrm{Min}}$ are the Abraham and Minkowski light momentums, respectively; $\mathbf{p}^{\mathrm{med}}_{\mathrm{kin}}$ is the medium kinetic momentum (also called ``Abraham material momentum'' in some literature \cite{r13}), and $\mathbf{p}^{\mathrm{med}}_{\mathrm{can}}$  is the medium canonical momentum (also called ``Minkowski material momentum'' \cite{r13}).  According to relativistic analysis of the Einstein-box thought experiment \cite{r21}, the total-momentum model should be modified into  $\mathbf{p}^{\mathrm{med}}_{\mathrm{kin}}+\mathbf{p}_{\mathrm{Min}}=$ total momentum, because the medium-box kinetic momentum--energy and Minkowski photon momentum--energy, respectively, constitute a Lorentz four-vector before and after the photon enters the box. In other words, in a system consisting of massive particles and photons, the momentums and energies of all individual massive particles and photons constitute Lorentz four-vectors no matter whether they have interactions or not.   \vspace{2mm} \\
\indent \textbf{(v) Apparent photon velocity and displacement.} We have shown that there may be \emph{apparent photon velocity} and \emph{apparent photon displacement} in a moving medium (see Fig. \ref{fig2}).  This conclusion comes from the fact that the photon propagation velocity is the phase velocity, for which there is no ``phase velocity four-vector'', and thus the photon does not have a four-velocity like a massive particle.  When using the time--space four-vector to describe a photon's motion, the space coordinates may not reflect its real location; thus resulting in the appearance of apparent velocity and displacement.  In addition, the fact that the photon does not have a four-velocity also calls into question the justification of the four-vector covariance imposed on the EM energy velocity in Laue--M\o ller theory because the photon is the carrier of EM energy.    \vspace{2mm} \\
\indent \textbf{(vi) Lorentz invariance of the Plank constant.}  We have shown that the Planck constant is a Lorentz invariant for a plane wave in a uniform medium (including empty space), which is a strict result of special relativity and the Einstein light-quantum hypothesis, while the Planck constant as a Lorentz invariant is an implicit postulate in Dirac relativistic quantum mechanics \cite{r40}.  This result will make the fine structure constant also a Lorentz invariant, which is explained as follows.  \\
\indent As shown in Sec.\ 11.\ 9 of the textbook by Jackson \cite{r25}, Maxwell equations $[\nabla\times\mathbf{H}-\partial(c\mathbf{D})/\partial(ct),\nabla\cdot(c\mathbf{D})]=(\mathbf{J},c\rho)$ and $[\nabla\times\mathbf{E}-\partial(-c\mathbf{B})/\partial(ct),\nabla\cdot(-c\mathbf{B})]=(\mathbf{0},0)$ can be written as $\partial_{\mu}G^{\mu\nu}(\mathbf{D},\mathbf{H})=J^{\nu}$ and $\partial_{\mu}\mathscr{F}^{\mu\nu}(\mathbf{B},\mathbf{E})=0$.  $G^{\mu\nu}(\mathbf{D},\mathbf{H})$ and $\mathscr{F}^{\mu\nu}(\mathbf{B},\mathbf{E})$ are assumed to be four-tensor Lorentz covariant to keep Maxwell equations invariant in form in all inertial frames, and thus $(\mathbf{J},c\rho)$ must be a four-vector. On the other hand, the electron's moving velocity must be less than light speed.  Accordingly, the electron charge must be a Lorentz invariant, as shown in Ref.\ \cite{r41}, although it is usually taken as an experimental invariant (see p. 555 of Ref. \cite{r25}, for example).  Now that light speed $c$, Planck constant $\hbar$, and electron charge $e$ are all Lorentz invariants, the fine structure constant $\alpha=e^2/\hbar c$  (in CGS unit) is also a Lorentz invariant. \\ 
\indent It should be emphasized that the proof of Lorentz invariance of the Planck constant $\hbar$ for a plane wave has great significance.  That is because, as mentioned previously, to meet Fermat's principle and global momentum--energy conservation law in the principle-of-relativity frame, the photon momentum and energy must constitute a Lorentz four-vector $P^{\mu}=\hbar K^{\mu}$, while the invariance of $\hbar$ is the sufficient and necessary condition for $\hbar K^{\mu}$ to be a four-vector. \vspace{2mm} \\
\indent \textbf{(vii) Classical mathematical conjecture.}  In relativistic dynamics, there is a well-known ``classical mathematical conjecture'', which states: 
\begin{quote}
If $\Theta^{\mu\nu}(X^{\sigma})$  is a Lorentz four-tensor defined on the domain $V(X^{\sigma})$, with $X^{\sigma}=(\mathbf{x},ct)$, and it is symmetric ($\Theta^{\mu\nu}=\Theta^{\nu\mu}$) and divergence-less ($\partial_{\mu}\Theta^{\mu\nu}=0 \Leftrightarrow \partial_{\nu}\Theta^{\mu\nu}=0$), then the time-row (column) integrals at any given time (assumed to be convergent) 
\begin{equation}
P^{\nu}=\int_V\Theta^{4\nu}d^{3}x \left( =\int_V\Theta^{\nu 4}d^{3}x \right)
\label{eq68}
\end{equation} 
constitute a Lorentz four-vector.  
\end{quote} 
This conjecture has been thought to be a well-established result of tensor calculus \cite{r42}, and it was used as a starting point in relativistic analysis of the Einstein-box thought experiment for resolution of the Abraham--Minkowski debate \cite{r43}.  However, we have shown in Ref.\ \cite{r41} that this conjecture is not true. \\
\indent Why is the conjecture not true?  As we know, the correctness of a mathematical conjecture cannot be legitimately affirmed by enumerating \emph{specific} examples, no matter how many; however, it can be directly negated by finding specific counterexamples, even only one.  As shown in Ref.\ \cite{r41}, the charged metal sphere is a specific counterexample to negate the conjecture, because the EM stress-energy four-tensor $T^{\mu\nu}$  for the charged metal sphere is symmetric ($T^{\mu\nu}=T^{\nu\mu}$) and divergence-less ($\partial_{\mu}T^{\mu\nu}=0$), but the time-row (column) integrals $\int_V T^{4\nu}d^3x ~(=\int_V T^{\nu 4}d^3x )$, or 
\begin{equation}
\left(\int_V \frac{\mathbf{E}\times\mathbf{H}}{c}d^3 x, \int_V \frac{1}{2}(\mathbf{D}\cdot\mathbf{E}+\mathbf{B}\cdot\mathbf{H})d^3 x\right)
\label{eq69}
\end{equation} 
never constitute a Lorentz four-vector.  In other words, the total (Abraham = Minkowski) EM momentum $\int_V[(\mathbf{E}\times\mathbf{H})/c^2]d^3x$ and energy $\int_V 0.5(\mathbf{D}\cdot\mathbf{E}+\mathbf{B}\cdot\mathbf{H})d^3x$    carried by the charged metal sphere cannot constitute a four-vector.  This conclusion is also clearly supported by the direct calculation result that, in an ideal planar-plate capacitor the total ``field's momentum--energy is seen to behave in a way that is not expected from a four vector'', as claimed by Mansuripur and Zakharian \cite{r18}.  Thus the ``classical mathematical conjecture'' is indeed not correct.    \vspace{2mm} \\
\indent \textbf{(viii) Intrinsic Lorentz violation.}  We have shown in Ref.\ \cite{r44} that there exists a new physics of so-called ``intrinsic Lorentz violation'' within the frame of the two postulates of special relativity (principle of relativity and constancy of light speed).  Traditionally, ``Lorentz invariance'' refers to all mathematical equations expressing the laws of nature must be invariant in form \emph{only under the Lorentz transformation}, and they must be Lorentz scalars, four-vectors, or four-tensors ... (see p. 540 of Ref. \cite{r25}, for example).  In other words, the Lorentz invariance is a single requirement that combines the two postulates together, and it is equivalent to the two postulates \cite{r45}.  Unfortunately, as shown in Ref.\ \cite{r44}, this is not true for the Doppler effect from a moving point light source, where the Doppler formula cannot be obtained from the Lorentz transformation but is exactly a result of the two postulates.  This phenomenon, called ``intrinsic Lorentz violation'', has never been realized in the community.  \\
\indent In fact, there is also an interesting ``intrinsic Lorentz violation'' for the plane wave in a moving uniform medium.  As shown in Sec.\ II herein, there are two velocities associated with the photon.  One is the photon velocity, namely, the phase velocity $\mb{\beta}_{ph}c=(\omega/|n_d\mathbf{k}|)\mathbf{\hat{n}}$, defined based on the wave four-vector $K^{\mu}=(n_d\mathbf{k},\omega/c)$, which is completely consistent with the two postulates.  The other is the apparent photon velocity $\mathbf{u}=(\mathbf{E}\times\mathbf{H})/W_{em}$, which is the EM ``energy velocity'' \cite{r16} or ``group velocity'' traditionally \cite{r30}.  The principle of relativity requires that a physical law be invariant in form and, of course, keep the same physical meaning in all inertial frames.  $\mb{\beta}_{ph}c$ and $\mathbf{u}$ are equal in the medium-rest frame, and they are invariant in form in any inertial frames.  $\mb{\beta}_{ph}c$  is the photon velocity in any inertial frames and it denotes the ``law of EM energy transport'', but $\gamma_{ph}(\mb{\beta}_{ph}c,c)$  is not a four-vector.  In contrast, $\gamma_{u}(\mathbf{u},c)$  is a four-vector and $\mathbf{u}=(\mathbf{E}\times\mathbf{H})/W_{em}$  is the photon velocity in the medium-rest frame, but under the four-vector Lorentz transformation of $\gamma_u (\mathbf{u},c)$, $\mathbf{u}$  is not the photon velocity any more in general (with physical meaning changed); accordingly, $\mathbf{u}=(\mathbf{E}\times\mathbf{H})/W_{em}$  cannot now denote the law of energy transport.  From this it follows that, (a) $\mb{\beta}_{ph}c$  denotes a physical law, but $\gamma_{ph}(\mb{\beta}_{ph}c,c)$  is not a four-vector; and (b) $\mathbf{u}$  does not denote a physical law, but $\gamma_u (\mathbf{u},c)$  is a four-vector.  Thus the two cases both break the definition of ``Lorentz invariance'', resulting in ``intrinsic Lorentz violation''.  (Note: $\gamma_{ph}(\mb{\beta}_{ph}c,c)$ and $\gamma_u (\mathbf{u},c)$  do not exist in free space where $|\mb{\beta}_{ph}c|=|\mathbf{u}|=c$ and both $\gamma_{ph}$ and $\gamma_{u}$ $\rightarrow\infty$.)  \\
\indent It should be indicated that the intrinsic Lorentz violation exposed in Ref.\,\cite{r44} and herein is essentially different from the ``Lorentz violation'' presented in Ref.\,\cite{r46}.  The intrinsic Lorentz violation takes place within the frame of the two postulates, and it is completely consistent with special relativity.  In contrast, the Lorentz violation \cite{r46} describes deviations from the two postulates; for example, there has been a controversy recently about whether there are deviations in the time dilation predicted by special relativity in experiments of high-energy ions \cite{r47}. \vspace{3mm} \\
\indent \textbf{(ix) Summary.}  In summary, in this paper we have answered the following most fundamental questions in classical physics and quantum physics.  When there exist dielectric materials in space:
\begin{itemize}
\item Is the principle of relativity still valid?
\item Are the Maxwell equations, momentum--energy conservation law, Fermat's principle, and Einstein light-quantum hypothesis equally valid in all inertial frames of reference?
\item Why is the EM momentum--energy stress tensor not enough to correctly define light momentum?
\item Why is the principle of relativity needed to identify the justification of the light-momentum definition?
\item Does the Poynting vector always represent EM power flow in any system of materials?
\item Is the Planck constant a Lorentz invariant?
\item Does the photon have a Lorentz four-velocity like a massive particle?
\item Why must the photon momentum and energy constitute a Lorentz four-vector?
\item Why is Lorentz invariance not equivalent to the two postulates of special relativity? 
\end{itemize}



\end{document}